\documentclass[12pt]{iopart}
\usepackage[square,numbers,sort&compress]{natbib}
\usepackage[colorlinks=false, pdfborder={0.5 0.5 0}]{hyperref}
\usepackage{enumitem}
\usepackage{breakurl}
\usepackage{graphicx}
\begin{document}

\title{Braess' paradox in the age of traffic information}

\author{S Bittihn and A Schadschneider}
\address{Institut f\"ur Theoretische Physik, Universit \"at zu K\"oln,
  50937 K\"oln, Germany}
\eads{\mailto{as@thp.uni-koeln.de}, \mailto{bittihn@thp.uni-koeln.de}}

\begin{abstract}
The Braess paradox describes the counterintuitive situation that the
addition of new roads to road networks can lead to higher travel
times for all network users. Recently we could show that user optima
leading to the paradox exist in networks of microscopic transport
models. We derived phase diagrams for two kinds of route choice
strategies that were externally tuned and applied by all network
users. Here we address the question whether these user optima are
still realized if intelligent route choice decisions are made based
upon two kinds of traffic information. We find that the paradox
still can occur if the drivers 1) make informed decisions based on
their own past experiences or 2) use traffic information similar to
that provided by modern navigation apps. This indicates that modern
traffic information systems are not able to resolve Braess' paradox.
\end{abstract}


\submitto{\JSTAT}

\vspace{2cm}

\begin{center}
  \today
\end{center}

\maketitle

\tableofcontents

\section{Introduction}

Everyday experience shows that we spend a lot of time in traffic jams
\cite{cookson2017inrix,tomtomtrafficindex,worldurbanizationprospects}. This
time can add up to more than 100h per year in certain parts of the
world \cite{duranton2011fundamental}.  Two potential solutions to the
problem of congestions come to mind: 1) building more roads and 2)
providing traffic information for the drivers to make better
decisions. However, it is known for some time that new roads are not
necessarily a solution.  The {\em Fundamental Law of Road Congestion}
\cite{duranton2011fundamental} states that more roads will lead to
more road users which will lead to congestions on the new roads as
well.

Even if the traffic volume does not increase, new roads will not
necessarily lead to an improvement, e.g. shorter travel times for the
drivers. This paradoxical fact has been established rigorously 50
years ago by Braess \cite{braess68,braessnw05} and is now commonly
known as {\em Braess' paradox}. In compact form it can be formulated
in the following way:
\begin{quote}
In road networks of selfish users additional roads
can lead to higher travel times for all users.
\end{quote}
Although it is formulated using terminology from traffic
engineering, the paradox has been shown to occur in a variety of
other systems as well, ranging from general transport networks
\cite{SteinbergZ83,DafermosN84,PasP97} to mechanical and electrical
systems \cite{PenchinaP03,cohen1991paradoxical}, pedestrian
dynamics~\cite{crociani2016}, microfluidic networks~\cite{case2019},
oscillator networks and power
grids~\cite{witthaut2012braess,tchuisseu2018}.
A review of some examples from mechanical systems, biological
networks, to power grids can be found in
\cite{motter2018antagonistic}.

Braess has exemplified the paradox for a simple network that has
just five edges. One of the essential ingredients for the occurrence
of the paradox is that one has to distinguish two different kinds of
optima in the system, the user optimum (uo, also called Nash
equilibrium), and the system optimum (so)~\cite{wardrop1952}.
The uo is realized if network users distribute themselves onto
the routes such that the travel times of all used routes are equal
and lower than those of any unused routes. It reflects the
perspective of individual drivers who can not improve their travel
times by choosing a different route and is thus widely considered to
be the stable state of a traffic network used by selfish users. The
so corresponds to a global perspective, e.g. of an engineer or
politician. In the so the drivers are distributed onto the routes
such that a global parameter is minimal. Several different global
parameters could be considered. Prominent examples are the weighted
average of all travel times~\cite{wardrop1952} or the sum of the
travel times of all drivers~\cite{stewart1980equilibrium}. Braess
considered yet another definition of the system optimum: in his
work~\cite{braess68,braessnw05} as well as in our previous works on
the Braess paradox~\cite{bittihn2016,bittihn2018} and also in the
present article the system optimum is defined as the state which
minimizes the maximum travel time of all used roads. Indeed, the
uo and the so can be different in certain situations,
i.e. correspond to a different distribution of drivers on the
available routes.

Braess' example \cite{braess68,braessnw05} consists of simple
networks with 4 and 5 links (where links represent roads),
respectively. In the network with 4 links, the uo and so
coincide. The addition of a new road (i.e. the fifth link) changes
the network such that the uo and so become different, or
-- more specifically -- in the network that includes the new road,
the individual travel times in the uo are larger than those in
the so. The travel time functions are chosen in such a way,
that the travel times in the uo of the network with the fifth
link are higher than those in the network without the fifth link.

Although the paradox considers a simplified scenario, it has been
observed in several real world situations. Newly built roads can
worsen the traffic situation, or inversely, the closing of roads can
improve traffic \cite{Kolata,baker2009, vidal2006}. A concept
related to Braess' paradox is the {\it price of anarchy}. It
measures the reduced efficiency of a system due to the selfish
behavior of agents \cite{YounGJ08}.

Braess' original work is based on a macroscopic mathematical model
of freeway traffic. A lot of further research effort was made and
let to a more coherent understanding of the paradox in such
mathematical
models~\cite{stewart1980equilibrium,murchland1970braess,frank1981braess,SteinbergZ83,DafermosN84,PasP97,nagurney2010negation}.
In previous works \cite{bittihn2016,bittihn2018} we have studied the
original Braess network with a more realistic traffic dynamics that
e.g. includes stochastic fluctuations. In these microscopic models,
all cars are considered individually. By varying the route choice
decisions of the drivers, the user optima of the networks with and
without the new road were found. The travel times in the user optima
of the two different versions of the networks were then compared.
These systems show rich phase diagrams which include extended
regions where Braess' paradox can be observed. More specifically,
two different cases of the drivers' route choice decisions have been
distinguished: 1) routes are chosen stochastically by each driver
\cite{bittihn2016}, and 2) drivers use fixed strategies for their
route choices \cite{bittihn2018}.

In the aforementioned research based on macroscopic and microscopic
models, Braess' paradox is found in the sense that, for the same
amount of cars, user optima exist such that those of the networks with
the new road have higher travel times than those of the corresponding
networks without the new road. For the case of perfectly rationally
deciding drivers with access to perfect traffic information, one can
assume that these user optima would indeed be realized and that thus
the Braess paradox would really be observed. The question that remains
is \emph{if those potentially accessible user optimum states are also
  realized in more realistic situations with real (imperfectly
  deciding) human network users who base their decisions on real
  (imperfect) traffic information.}

In fact it has been shown that often travel time minimization is not
the only factor determining route choice
decisions~\cite{parthasarathi2013,chen2001using,zhu2015} and that even
if it is, drivers do not decide perfectly rational on this
basis~\cite{rapoport2009choice,selten2007}. Therefore variations of
the above mentioned definition of the user optimum were
introduced~\cite{daganzo1977stochastic,mahmassani1987boundedly}. While
perfect traffic information is also not present in road networks, with
the introduction of smartphone routing apps and personal navigational
systems, more accurate information is
available~\cite{meneguzzer2013day}. It has recently been shown, that
this might actually lead to the realisation of user optima in some
cases of road networks~\cite{madrigal2018,cabannes2018impact}.

Here we examine whether Braess' paradox is realized in a microscopic
transport model if users decide intelligently based upon information
similar to that available for real modern day road
networks. This is meant in the following sense: are the user optima
in the systems with and without the new road which are accessible by
externally tuning the users' decisions reached if users
intelligently base their decisions on information similar to that
available in present day real world networks. Two types of traffic
information are considered: information based on the drivers own
memory or experience (as e.g. in a commuter scenario) and
information similar to that provided by smartphone apps. It is shown
that both types drive the system into its user optima, realizing
Braess' paradox. This is a strong indication for answering the
question if the paradox still occurs in present day road networks or
if it is potentially resolved due to modern traffic information: it
seems that the paradox is indeed still of great importance!  It is
furthermore shown that user optima of different phases of the system
(that do not show Braess' paradox) are also realized. We conclude
that the phase diagrams derived in~\cite{bittihn2016,bittihn2018}
are `realistic'.

\section{Background information}

Before presenting the results of our study we define some
terminology and provide some more details about Braess' paradox and
some background to traffic information and route choice processes.
Then we define the model investigated here and give a short summary
of previous results.

\subsection{Some terminology}

Since this paper is concerned with traffic networks we will use the
terms ``edge'' and ``road'' interchangeably.  A ``route'' is a
connection between an origin and a destination in a traffic network.
A route can be comprised of multiple roads and also of so-called
``junction sites'' which connect roads. The ``travel time'' of a
road refers to the time it takes to traverse the road. The travel
time on a route refers to the time needed to traverse the route,
i.e. to get from the origin to the destination on that route.
Furthermore, the terms ``car'' and ``particle'' are used
interchangeably as well as ``driver'', ``user'' and ``agent''.

A ``strategy'' of an agent refers to its route choice. Two specific
types of strategies that were used in our previous
research~\cite{bittihn2016,bittihn2018} are ``pure strategies'' and
``mixed strategies''. For pure strategies the driver chooses exactly
one specific route, whereas ``mixed strategy'' refers to the case in
which one route is chosen out of several routes with a certain
probability. In real road networks, if a network user has to perform
route choices repeatedly, e.g. mixtures of these two strategy-types
can be at play. A ``state'' of the network is given by the
distribution of the cars onto the routes, i.e. the set of the
strategies of all drivers.

The ``user optimum'' state is often considered to be the stable
state of traffic networks with ``selfish users'', i.e. agents who
choose their routes non-altruistically only according to their own
intentions. The user optimum is reached if the travel times of all
cars are such that they are equal on all used routes
and, at the same time, lower than those
of any unused routes~\cite{wardrop1952}. A ``pure user optimum'' (puo) is
reached if all agents follow pure strategies. In this case the
numbers of cars using each route are integer numbers (or zero). This
corresponds to a ``pure Nash equilibrium'' in game
theory~\cite{nash1950}. A ``mixed user optimum'' (muo) is
reached if all agents follow mixed strategies and the \textit{mean}
values of the travel times of all used routes are equal and lower
than those of any unused routes. In this case the average numbers of
cars following routes can be positive non-integer values or zero. It
corresponds to a ``mixed Nash equilibrium''~\cite{nash1950}.

\subsection{The Braess paradox}
\label{sec:braess_paradox}

\paragraph*{The original example.}

\begin{figure}[ht]
  \centering
  \includegraphics[width=0.3\textwidth]{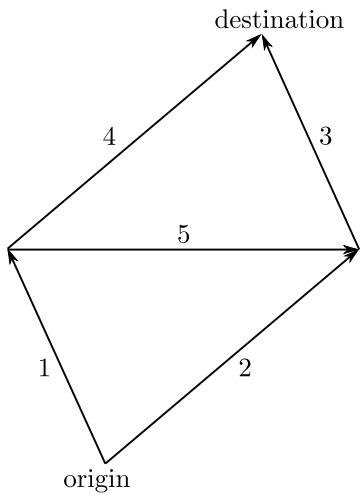}
  \caption{\label{fig:braess_networks}Braess' network as presented in
    his original work~\cite{braess68,braessnw05}. All
    agents move from the same origin to the same
    destination. There are three routes available, named
    after the edges they are comprised of: routes 14, 23 and 153. Edge
    5 is supposed to be newly added to the network. Route 153 is only
    available after this addition. The networks without and with edge
    5 are called 4link and 5link networks, respectively.}
\end{figure}

The network proposed by Braess in his original work
\cite{braess68,braessnw05} is shown in
Fig.~\ref{fig:braess_networks}. In his scenario all agents move from
the same origin to the same destination. Road 5 is the road which is
added to the system. Thus in the network without road 5, which we
call ``4link network'' from now on, there are two routes from origin
to destination: route 14 and route 23. In the network with road 5,
called ``5link network'' in the following, there is the additional
route 153 to the destination\footnote{Routes are named
according to the (ordered!) edges they are comprised of. From here
on, variables corresponding to the 4link and 5link networks are
marked with superscripts $(4)$ and $(5)$, respectively.}.
In the original example, traffic flow was characterized only by travel
time functions $T_i$ linear in the number of cars $n$ using road $i$:
\begin{eqnarray}
 T_1=T_3&=10n, \\
 T_2=T_4&=50+n, \\
 T_5&=10+n.
\end{eqnarray}
The 4link network is thus symmetric.

Braess showed that for a total number of $N=6$ cars, the pure user
optimum of the 4link network is given by half the cars using route 14
and the other half route 23, respectively:
$n^{(4)}_{14, \mathrm{puo}}=n^{(4)}_{23, \mathrm{puo}}=3$. This
results in equal travel times of both routes:
$T^{(4)}_{\mathrm{puo}}=T^{(4)}_{14, \mathrm{puo}}=T^{(4)}_{23,
  \mathrm{puo}}=83$. In the 5link network the pure user optimum is
given by
$n^{(5)}_{14, \mathrm{puo}}=n^{(5)}_{23, \mathrm{puo}}=n^{(5)}_{153,
  \mathrm{puo}}=2$ with equal travel times on all routes,
$T^{(5)}_{\mathrm{puo}}=T^{(5)}_{14, \mathrm{puo}}=T^{(5)}_{23,
  \mathrm{puo}}=T^{(5)}_{153, \mathrm{puo}}=92$.  Thus,
$T^{(5)}_{\mathrm{puo}}>T^{(4)}_{\mathrm{puo}}$, i.e. the new road
leads to larger pure user optimum travel times.

The paradox is also observed if all cars follow mixed
strategies\footnote{For a more detailed discussion of the mixed user
  optima, see~\ref{sec:app_muo}.}. For $N=6$, in the 4link network the
mixed user optimum state is found if routes 14 and 23 are chosen
with equal probabilities:
$p^{(4)}_{14, \mathrm{muo}}=p^{(4)}_{23, \mathrm{muo}}=1/2$. This
leads to equal travel time expectation values of
$\langle\, T^{(4)}_{\mathrm{muo}}\, \rangle=\langle\, T^{(4)}_{14,
  \mathrm{muo}}\, \rangle=\langle\, T^{(4)}_{23, \mathrm{muo}}\,
\rangle=88.5$.  In the 5link system the mixed user optimum is given
for $p^{(5)}_{14, \mathrm{muo}}=p^{(5)}_{23, \mathrm{muo}}=5/13$ and
$p^{(5)}_{153, \mathrm{muo}}=3/13$ with travel time expectation values
$\langle\, T^{(5)}_{\mathrm{muo}}\, \rangle=\langle\, T^{(5)}_{14,
  \mathrm{muo}}\, \rangle=\langle\, T^{(5)}_{23, \mathrm{muo}}\,
\rangle=\langle\, T^{(5)}_{153, \mathrm{muo}}\, \rangle=93.6923$. Thus
also in the case of mixed user optima,
$\langle\, T^{(5)}_{\mathrm{muo}}\, \rangle>\langle\,
T^{(4)}_{\mathrm{muo}}\, \rangle$.

\paragraph*{Results of further research.}

Since the initial description of the paradox by Braess a lot of efforts were made
to understand the phenomenon in more detail in the context of mathematical models of
traffic networks. It was shown that in the original model of Braess, the paradox occurs
for several amounts of total users, and not only for Braess' specific example of
$N=6$~\cite{PasP97}. The paradox also occurs for different choices of (linear) travel
time functions in Braess' original network~\cite{murchland1970braess} and also in different
network topologies~\cite{stewart1980equilibrium}. A general framework for predicting the
occurrence of the paradox in networks of uncorrelated links was
established~\cite{frank1981braess,SteinbergZ83}. Mathematical models including
correlations between the roads were studied in~\cite{DafermosN84}. It was furthermore
shown for arbitrary networks and models with monotonically increasing travel time
functions, that if the paradox occurs at a certain density, the new road will be ignored
completely for densities higher than a certain threshold~\cite{nagurney2010negation}.

\subsection{Braess' paradox in TASEP networks}

In an attempt to get an understanding of the paradox in a more
realistic context, in two recent
articles~\cite{bittihn2016,bittihn2018} we have shown that the
Braess paradox can also be observed in networks of stochastic,
microscopic traffic models, i.e. in networks of totally asymmetric
exclusion processes (TASEPs). The description of traffic flow in
Braess' original example, as summarized above, was rather basic,
being more of a proof of principle instead of a realistic model.
Braess used only linear travel time functions which is not
realistic. In addition, microscopic interactions and the stochastic
nature of traffic were omitted. Furthermore, correlations between
the roads were not taken into account.

Modelling traffic flow in the network by coupled TASEP segments is a
first step to a more realistic traffic description by including
these aspects while, at the same time, keeping the system simple
enough to be analysed. There is a vast amount of research dedicated
to the many variants of TASEPs and their properties. For some
condensed information the reader is referred to
e.g.~\cite{Schuetz-review,BlytheE07,schadschneider2010stochastic}.

\begin{figure}[ht]
  \centering
  \includegraphics{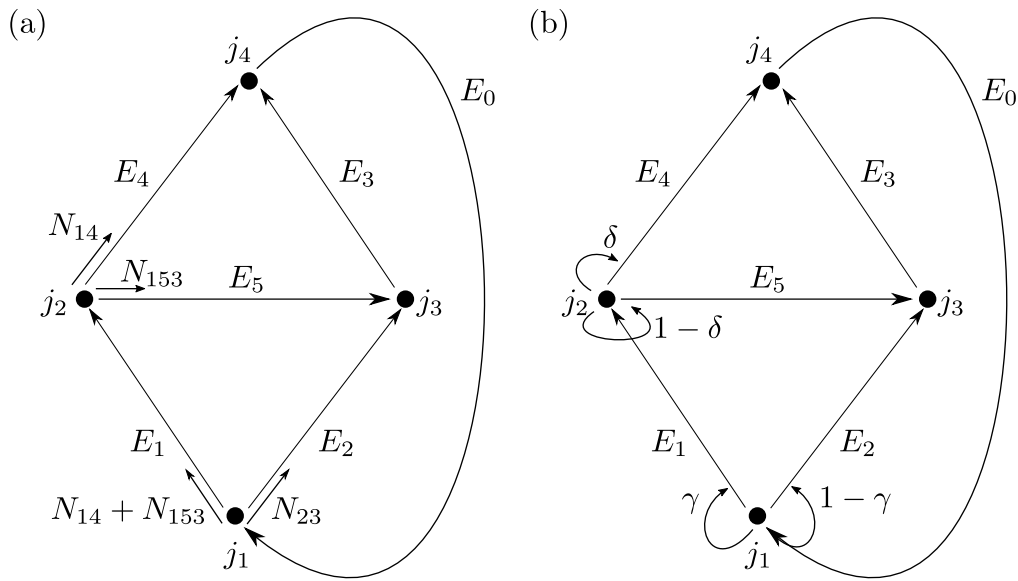}
  \caption{\label{fig:braess_networks_of_tasep}Braess' network of
    TASEPs with periodic boundary conditions for two different types
    of externally tuned strategies. The structure of the network
    corresponds to that used by Braess in his original article, as
    shown in Figure~\ref{fig:braess_networks}. Here, edges $E_1$ to
    $E_5$ are made up of TASEPs, coupled through junction sites $j_1$
    to $j_4$. The added edge $E_0$ realizes the periodic boundary
    conditions. Part (a) shows the system with fixed strategies, in
    which fixed numbers of cars $N_i$ use routes $i$ as studied
    in~\citep{bittihn2018}. In part (b) the system with turning
    probabilities, as studied in~\citep{bittihn2016}, is
    shown. Particles sitting on junctions $j_1$ or $j_2$ turn left
    with probabilities $\gamma$ and $\delta$, respectively.}
\end{figure}

The Braess network of TASEPs for two different route assignment
types is shown in Figure~\ref{fig:braess_networks_of_tasep}.
The network has the same structure as Braess' original network
(Fig.~\ref{fig:braess_networks}). The edges $E_i$ are now made up of
TASEPs of lengths $L_i$ joined through junction sites $j_k$.
Furthermore we use periodic boundary conditions via the additional
link $E_0$. This has the advantage that the total number of
particles $M$ is conserved which allows to compare the travel times
of the 4link and 5link systems in their respective user optima for
the same number of particles. To reduce the number of parameters, in
the following we will present results only for the following edge
lengths:
\begin{eqnarray}
L_1=L_3=100\,,\quad L_2=L_4=500\,,\quad L_0=1\,.
\label{eq:edgelenghts}
\end{eqnarray}
The length of $E_5$ will be varied, subject to the condition that the
length of the new route is smaller or equal to that of the two old
routes (which are of equal length).

Figure~\ref{fig:braess_networks_of_tasep}~(a) shows the network with
fixed personal strategies as analysed in~\cite{bittihn2018}. In this
case, each particle keeps its personal pure strategy of always
choosing one specific route. Numbers of $N_{14},~N_{23},$ and
$N_{153}$ particles choose routes 14, 23 and 153 respectively, with
$N_{14}+N_{23}+N_{153}=M$. The three numbers $N_{14},~N_{23},$ and
$N_{153}$ can also be expressed through the two quantities
\begin{eqnarray}
 n_{\mathrm{l}}^{(j_1)}&=1-\frac{N_{23}}{M}, \label{eq:nlj1}\\
 n_{\mathrm{l}}^{(j_2)}&=\frac{N_{14}}{N_{14}+N_{153}},\label{eq:nlj2}
\end{eqnarray}
which describe the fraction of particles turning 'left' at junctions
$j_1$ and $j_2$. User optima can be found by varying the
$N_{14},~N_{23},$ $N_{153}$. The user optima obtained in this
scenario are pure user optima.

In Figure~\ref{fig:braess_networks_of_tasep}~(b), the network with
the route assignment procedure governed by turning probabilities, as
studied in~\cite{bittihn2016}, is shown. In this case all particles
are equal. A particle on junction $j_1$ jumps to the left (i.e. onto
$E_1$) with probability $\gamma$ and to the right (i.e. onto $E_2$)
with probability $1-\gamma$. In the 5link network, particles on
junction $j_2$ jump left (i.e. onto $E_4$) and right (i.e. onto
$E_5$) with probabilities $\delta$ and $1-\delta$, respectively.
User optima in this network are found by varying the $\gamma$ and
$\delta$. They are mixed user optima.

These two types of route assignment are from here on called
``externally tuned strategies". This is meant in the sense that all
decisions are set at the beginning of each simulation run
\textit{externally}, i.e. not intelligently by the particles
themselves. These route choices will later on be distinguished from
route choice decisions made by `intelligent' agents following our
route choice algorithm.

\begin{figure}[ht]
  \centering
  \includegraphics{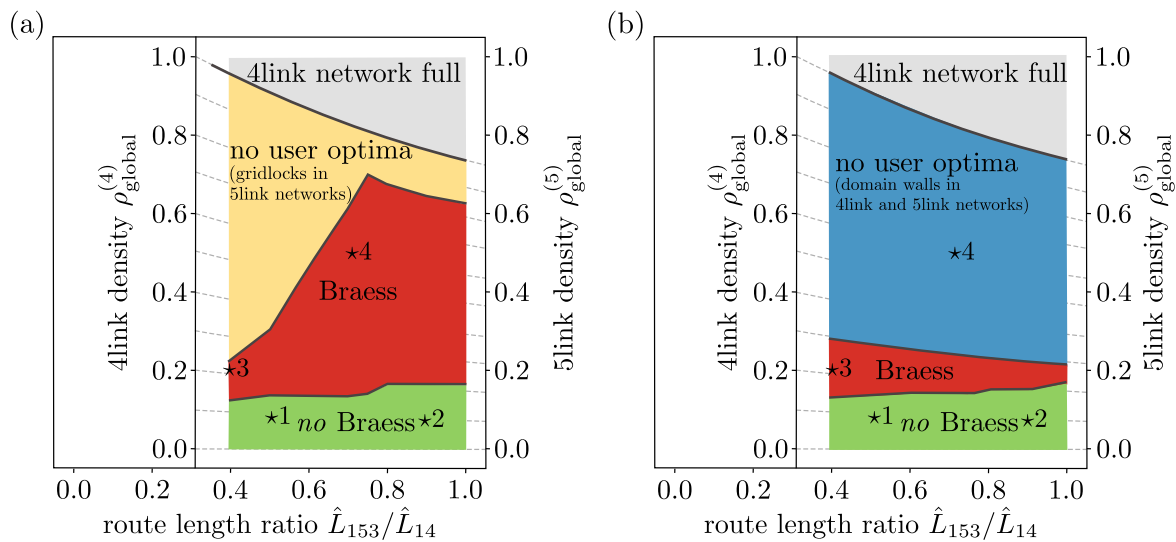}
  \caption{\label{fig:phase_diagrams}Phase diagrams of Braess' network
    of TASEPs (Fig.~\ref{fig:braess_networks_of_tasep}) for (a) fixed
    route choices and (b) turning probabilities. The phase of the
    system depends on the route length ratio
    $\hat{L}_{153}/ \hat{L}_{14}$ between the new and old routes and
    the global densities $\rho^{(4/5)}_{\mathrm{global}}$ in
    4link/5link systems. In both phase diagrams four states $\star 1$
    to $\star 4$ are marked. The route choice algorithm presented in
    the present paper is tested on these states
    (Sec.~\ref{sec:results}). The detailed parameters of the four
    states are described in~\ref{sec:app_test_states}. The shown phase
    diagrams are simplified, as e.g. sub-phases are not shown. For
    full details, see \citep{bittihn2018,bittihn2016}.
    }
\end{figure}

The phase diagrams for both networks are shown in
Fig.~\ref{fig:phase_diagrams}. The phase of the system depends on
the ratio $\hat{L}_{153}/ \hat{L}_{14}$ of the lengths of the new
route 153, $\hat{L}_{153}$, and the two old routes,
$\hat{L}_{14}=\hat{L}_{23}$; and the global density. Since the phase
of the system describes how the travel times of the 4link and 5link
systems' user optima are related, the global densities of both
systems $\rho^{(4)}_{\mathrm{global}}=M/(4+\sum_{i=0}^4 L_i)$ and
$\rho^{(5)}_{\mathrm{global}}=M/(4+\sum_{i=0}^5 L_i)$ are shown.

In both systems at low densities the user optima in the 5link networks
have lower travel times than those of the corresponding 4link
networks, thus the system is not in a Braess phase (``no Braess"/green
phases). The Braess paradox is observed in significant parts of the
phase spaces of both systems (``Braess"/red phases). In both systems
at higher densities no user optima could be found due to gridlocks
(orange phase) and fluctuating domain walls (blue phase) in the system
with fixed strategies and turning probabilities, respectively. For
more detailed phase diagrams and explanations, the reader is referred
to~\cite{bittihn2018,bittihn2016}.

The phase diagrams were obtained by assigning the particles onto
their routes externally via the $(N_{14},N_{23},N_{153})$ and
$(\gamma,\delta)$. By varying these parameters, user optima were
found and the travel times of the 4link and corresponding 5link
systems were compared to construct the phase diagrams. In the
present article we address the question if these user optima are
actually realized if the particles choose their routes
intelligently, based on different kinds of information. As in the
original example presented by Braess in his mathematical model, the
pure and mixed user optima for one combination of the
$(\hat{L}_{153}/\hat{L}_{14}, \rho^{(4/5)}_{\mathrm{global}})$ do
not necessarily coincide, i.e. the
$(n_{\mathrm{l}}^{(j_1)},n_{\mathrm{l}}^{(j_2)})$ and
$(\gamma,\delta)$ which realize a pure and mixed user optimum
respectively do not always have equal values.

In the following, we summarize some results of previous research on
traffic information and route choice processes and then present our
route choice algorithm. This algorithm is then applied to the four
states marked in Figure~\ref{fig:phase_diagrams}. The travel times
of the user optima of these states can be found
in~\ref{sec:app_test_states}.

\subsection{Types of traffic information}
\label{sec:types_traffic_info}

Information about the state of traffic networks is called {\it
traffic information}. It can consist of information on various
aspects of the traffic network, such as the positions of all
vehicles, average speeds, traffic light phases and many more. Here
we focus specifically on information about travel times on roads and
routes in the network.

Information available to network users can be grouped into two main
categories.
\begin{enumerate}
 \item ``Public Information" is in principle accessible to all network users.
 \item ``Personal Information" is only known to individual network
 users. It is usually based on the user's personal experience and/or
 specifically designed for a specific user, e.g.\ based on his/her current position,
 destination etc.
\end{enumerate}
These two main categories can contain information from three different
sub-categories~\cite{benakiva1991}.
\begin{enumerate}[label=\alph*)]
\item ``Historical Information" describes travel times measured in the
  network in previous time periods.
 \item ``Current Information" refers to the most up-to-date
   information available. It can be given in the form of providing
   network users with the current state of the network, e.g. providing
   the current traffic densities or the currently measured (average)
   speeds on certain routes as e.g. in \cite{VerkehrNRW}.

   If one sticks strictly to this definition, in real traffic networks
   travel time information cannot be current information. This is due
   to the following problem: if e.g. a network user finishes a trip in
   a given moment and her experienced travel time information is
   immediately made available to the public, this information does not
   represent the travel time in the network right in that moment. It
   is instead the travel time of the used route at the current time
   minus the measured travel time. In the current moment the traffic
   situation might have changed and a user choosing the same route
   right now might experience a different travel time.
 \item ``Predictive Information" is e.g. given in the form of
  estimated travel times for the
   routes. In contrast to the two other types of information,
   predictive information can -- by its nature -- not be guaranteed to be
   accurate. If predictive information is given to network users in
   the context of route choices, a specific dilemma occurs: the
   information potentially influences the agents, leading them to
   take certain route choice decisions which then change the traffic
   state and thus invalidate the information~\cite{wahle2000decision}.
   A special kind of predictive information is ``prescriptive
   information" which -- opposed to ``descriptive information" which
   only describes network states -- advises network users to use
   specific routes~\cite{meneguzzer2013day}.
\end{enumerate}
Network users may have access to combinations of all these types of
information and make their own individual route choice decisions based
upon them.

\subsubsection{Traffic Information Available in Present Day Road Networks}

In present day road networks various kinds of information are
available. Individual network users might have personal historical
information about travel times on specific routes based on their own
experiences if they used these routes before (personal historical
information, in this case also called ``experiential
information"~\cite{meneguzzer2013day}). This is often the case in
commuter scenarios. Furthermore, individual users might have some
insights from friends or other personal sources. Public information
of various kinds is available from numerous sources. Public
historical information can be found in internet databases and is
also available in various smartphone routing apps and personal
navigational systems. Current public information is available from
radio traffic forecasting and various advanced traffic information
systems (ATIS)~\cite{atis-wikipedia} such as variable road signs.

Personal navigational systems and smartphone routing apps also
provide predictive public information. This type of information is
considered public since these devices are in principle accessible
for everyone. In contrast to the former mentioned information, these
tools also provide prescriptive route choice information. Among many
alternatives, Google Maps was the most popular routing app in the US
in 2018~\cite{statista-maps}. The main difference from more
traditional types of traffic information is that such apps rely on
crowdsourcing~\cite{google-blog-maps}. This means that all users of
the app send their location data to Google where this data is in
turn used to get an accurate picture of the current traffic
situation of the network (given there are enough Google Maps users
in the network at that moment). This current information is combined
with large quantities of historical information to provide fairly
accurate public predictive (prescriptive) information (about travel
times). Details about how the Google Maps algorithm works are not
known to the public~\cite{google-blog-maps,how-google}.

To approximate the situation which is found in real modern road
networks in this article we consider user decisions based on personal
historical and public predictive information. A mixture of these two
types of information is found in real road networks, especially in
commuter scenarios.

The detailed knowledge of current traffic states available due to
crowd sourced information could also be used to provide information
to road users that aims at optimizing the state of the {\it whole}
traffic system. Such information would be given in different forms
than just information about (predictive) travel times on available
routes. It could be designed e.g. to drive traffic networks into
their system optima and could thus lead to a reduction of traffic
congestion, as shown e.g. in~\cite{li2020reducing}. Even though the
necessary data for such information is in principle available, such
systems are (to our knowledge) currently not in use, which is why
they are not considered here.

\subsection{Previous research on route choices}

The question how road users choose their routes given certain types of
traffic information and consequently the question if user optima are
realized in networks of selfish users has been analysed in various
scientific disciplines. The research approaches can be subdivided into
three groups: analyses of real world data, mathematical models and simulation
studies, and laboratory experiments.

\subsubsection{Analysis of real world data}
Since traffic networks are generally highly complex structures with
numerous users that all decide individually it is difficult to gain
objective knowledge about what drives the decision making processes
in these networks. Next to much anecdotal knowledge about route
choices, some large scale real world observations and experiments
were performed. A study from 2001~\cite{chen2001using} hints at
travel time minimization not being the only factor driving route
choice decisions. Furthermore, travel time seems to be
systematically misperceived by many
drivers~\cite{parthasarathi2013}. In a study in which vehicles of a
large number of network users were equipped with GPS units it was
shown that only approximately one third of traffic network users
chooses the fastest available path~\cite{zhu2015}. For a nice recap
of results of previous research, the reader is also referred
to~\cite{zhu2015}.

The introduction of smartphone routing apps lead to a different
usage of road networks in many parts of the world. Data suggests
that minimal travel time becomes more important when using those
apps and hints at the realization of user optima due to the heavy
usage of these tools~\cite{cabannes2018impact}. There are also many
negative side-effects of such tools, such as increasing use of
smaller side roads to avoid congested main roads, leading to
complaints by residents~\cite{thai2016negative}.

An effect that has already been predicted in mathematical traffic
models~\cite{horowitz1983, benakiva1991} and queueing
models~\cite{hall1996} and was observed in
simulations~\cite{wahle2000decision} has also been observed as a
consequence of routing apps: routing apps provide predictive
information about travel times based on the current network state to
all users. This information may then influence the users' decisions
in such a way that these decisions invalidate the prediction. This so-called
``overreaction"~\cite{benakiva1991} has been observed in networks with
routing apps~\cite{cabannes2018impact}.

\subsubsection{Mathematical models and simulation studies}

In earlier days, research on route choices was performed using
mathematical models~\cite{benakiva1991,horowitz1983,arnott1991}
which already predicted many effects that were later confirmed e.g.
by simulations. In~\cite{bazzan2003learning,bazzan2005case} the
Braess paradox was considered in a discrete time macroscopic
mathematical model. It was shown that in the given model, the user
optima leading to Braess' paradox can be avoided with a special kind
of personal historical information.

For route choice research on the basis of {\em microscopic}
simulations it has proven to be useful to implement so-called
``multi-agent techniques"~\cite{bazzan1999agents,wahle2000decision}.
The traffic flow itself, forming the ``tactical layer", is modelled
by a stochastic microscopic traffic model. The acquisition of
information and the route choice decisions, the so-called
``strategic layer", are modelled by an algorithm. This multi-agent
approach is also used in the present paper.

Multi-agent models with the tactical layer being described by the
Nagel-Schreckenberg model~\cite{nagel1992} have been studied for
certain types of traffic information: It was shown that the
availability of public historical information in a symmetric
two-route network with open boundary conditions leads to
oscillations around the user optimum~\cite{wahle2000decision}.
In~\cite{wahle2000decision} the latest experienced travel times on
both routes were made available to all network users. This leads to
overreactions since this travel time information is based on the
network states previous to when the information is made available.
The user optimum is reached when
agents choose each route with equal probability. Instead
oscillations between periods of all cars using just one route and
times of all cars using just the other route are realized. This
observation lead to the proposition of many other types of
information with the aim of realizing the user optima in this
specific network (see e.g.~\cite{lee2001effects,wang2005,chen2012}).
A good review of these information types and how they perform is
found in~\cite{he2014}. There it is also pointed out that most of
these realize user optima only in the specific symmetric two route
scenario. It was also shown that the paradox can be observed in the
Braess network, if traffic flow is modelled by the
Nagel-Schreckenberg model~\cite{bazzan2005case}.

In~\cite{levy2016emergence} a two route model with periodic boundary
conditions and dynamics similar to TASEP is studied with users with
personal historical information, i.e. users that have memories of
certain lengths that decide based on their own experiences. It is
shown that this type of information realizes user optima in the
network. This is similar to some of the results to be presented
later in the present paper.

Large-scale simulations of systems with information similar to that
provided by smartphones suggest that user optima are realized in
these systems~\cite{cabannes2018impact}. To our knowledge no models
based on simulations in connection with personal historical
information or public predictive information like that provided by
smartphones, implemented in the way given in the present paper, have
been studied in controlled small networks like the present Braess
network.

\subsubsection{Laboratory experiments}
Next to research based on mathematical models and simulations which
was mainly conducted in the traffic science and traffic engineering
community, in the social sciences and behavioral economics many
laboratory experiments on route choice processes were performed with
the aim of understanding how humans decide. Typically, a road
network is implemented and the traffic flow is described by a
mathematical model. Human subjects are then asked to perform route
choices repeatedly given various types of information. Usually real
money is paid out as an incentive to perform well in the task of
travel time minimization.

A nice review of route choice experiments is found
in~\cite{ben2015response}. Here we focus on experiments that either
directly address the Braess paradox or are closely
related. In~\cite{selten2007} and~\cite{meneguzzer2013day} scenarios
with two and three unconnected links from origin to destination were
studied, respectively. In both studies, when participants relied on
personal historical information, i.e. their own experience of travel
times of previous rounds, user optima were reached on average with
some persisting fluctuations. In~\cite{selten2007} also the situation
with public historical information, in which participants had
knowledge of travel times also on routes not taken, was studied. The
user optimum was also reached on average.

In~\cite{ye2017} route choice decisions in the 5link version of the
Braess network were studied in a virtual experiment. Participants had
to chose routes daily in the app ``WeChat". Subject to public
historical information the user optimum was reached.

In~\cite{rapoport2009choice} the Braess paradox was tested directly
in the laboratory. Participants performed route choices first in
Braess' 4link and then in the 5link network. Subject to public
historical information user optima were reached in both cases (in
the 4link on average, while fluctuations decreased in the 5link) and
the paradox was realized. In a further network of different
topology, Braess' paradox was realized as
well~\cite{rapoport2009choice}. Furthermore, in~\cite{mak2018braess}
another network exhibiting Braess-behaviour was studied. The paradox
was also realized here.

\subsubsection{How research from the different areas connects}
The research performed in the different scientific disciplines
employing the various approaches mentioned above all add to the
understanding of route choice processes while all of them have their
advantages and disadvantages. In observations of real world data,
typically the system cannot be controlled as well as in the toy
systems studied in simulations and laboratory experiments. Here,
there is always a larger underlying network. The objective of
the `participants' is not clear either. Nevertheless, on one hand,
this lead to important discoveries such as the minimization of
travel time not necessarily being the sole goal of network users. On
the other hand, these conclusions cannot be proven rigorously, since the objectives
and motivations for the route choice of the agents are not known.

The two more controlled approaches also differ in important ways: in
all the works cited in the paragraph on laboratory experiments, the
traffic description is limited to deterministic, macroscopic travel
time functions of the individual roads in the network. Microscopic
interactions are omitted and furthermore correlations between the
roads are not modelled. Since these omitted details lead to a much
more complicated traffic behavior in real traffic networks, as
travel times could e.g. change more drastically if many drivers
change their strategies, one can argue that drivers reactions could
also be different in these cases. Thus it is not really clear if the
laboratory observations really transfer to the real world. In
research based on simulations, the tactical layer, i.e. the
description of traffic flow, is more realistic, whereas the route
choice decisions, i.e. the strategic layer, is not as realistic
since it is not done by real humans.

Drawing from the observations of real world data one also has to note
that the results from laboratory experiments cannot be taken as facts
since in real world data it was observed that travel time minimization
is not necessarily the main objective of drivers. In laboratory
experiments nevertheless it is the sole objective by design of the
experiment.

\section{Route choice algorithms}
\label{sec:route_choice_alg}

To find out whether user optima are realized in the 4link and 5link
versions of a Braess network of TASEPs
(Fig.~\ref{fig:braess_networks_of_tasep}), when used by
intelligently deciding agents, we implemented the following route
choice algorithm. We examined variants where the agents have access
to personal historical and public predictive information. With this
we want to combine a more realistic tactical layer (i.e. a
microscopic traffic modelled by coupled TASEPs) with a more
realistic strategic layer with decisions based on realistic traffic
information.

Due to the periodic boundary conditions all $M$ particles stay in
the system and thus decisions based on personal experiences, i.e. on
the memory of the particles, can be implemented.

The system is always initialized by placing $M$ particles on the
routes randomly and assigning an initial pure strategy (to choose
either route 14, 23 or 153) randomly to each agent. The system then
undergoes a relaxation procedure which is different depending on the
type of information used. The relaxation procedures are explained in
the following paragraphs which describe the different types of
implemented information. Once the system is relaxed, all agents have
information about the (expected) travel times of all routes. The
information for all three routes 14, 23 and 153 is from now on
called $T_{14,\mathrm{info}}$, $T_{23,\mathrm{info}}$ and
$T_{153,\mathrm{info}}$, respectively. If an agent went once from
$j_1$ to $E_0$ (i.e. after jumping out of $j_4$), this agent has
finished one ``round".

After relaxation is complete, route choice decisions can occur at
three points: before starting a new round (when `sitting' on $E_0$)
and during a round when sitting on $j_1$ or $j_2$ and not being able
to jump to the desired target site.  Before a new round, when sitting
on $E_0$ before jumping to $j_1$, each particle generally chooses the
route $i$ with the lowest $T_{i,\mathrm{info}}$. To make such
decisions more realistic, the two following parameters are introduced
to the algorithm.
\begin{enumerate}
\item With probability $p_{\mathrm{info}}$ the particle bases its
  decision on the available $T_{i,\mathrm{info}}$. With probability
  $1-p_{\mathrm{info}}$ the particle chooses one of the two or three
  routes (depending on whether the 4link or 5link system is simulated)
  randomly. The random decisions are introduced to account for the
  findings from observations in the real world, that users do not
  wholly base their decisions on the objective of minimizing travel
  times.
 \item If, with probability $p_{\mathrm{info}}$ as described in (i),
   an information-based decision will be taken, the difference between
   the expected travel times on the routes
   $\Delta T = |T_{14,\mathrm{info}}-T_{23,\mathrm{info}}|+|T_{14,\mathrm{info}}
   -T_{153,\mathrm{info}}|+|T_{23,\mathrm{info}}-T_{153,\mathrm{info}}|$
   is calculated\footnote{In the 4link system this expression reduces
     to $\Delta T=|T_{14,\mathrm{info}}-T_{23,\mathrm{info}}|$}. If
   this difference is below the threshold of
   $\Delta T_{\mathrm{thres}}$, the agent stays on the route of
   the previous round. If $\Delta T\geq \Delta T_{\mathrm{thres}}$,
   the agent switches to the route with the lowest
   $T_{\mathrm{info}}$. Thus the agents act ``boundedly
   rational"~\cite{mahmassani1987boundedly}.
\end{enumerate}

Additionally to these decisions \textit{before} any new round,
agents can make route choice decisions \textit{during} the rounds.
These decisions work as follows. Consider an agent in the 4link
network sitting on $j_1$ who chose to take route 23 before the round
began. If this agent cannot jump to its target site (first site of
$E_2$) since this site is occupied, (s)he may re-decide for another
route. If $T_{23,\mathrm{info}}\geq T_{14,\mathrm{info}}$ (agent
chose route 23 based on a random decision before the round), (s)he
will then immediately switch to route 14. If
$T_{23,\mathrm{info}}<T_{14,\mathrm{info}}$, the particle will keep
trying to jump onto $E_2$ for $\kappa_{j_1,\mathrm{thres}}$ times
the to-be-expected saved time on route 23, i.e. for
$\kappa_{j_1,\mathrm{thres}}\cdot(T_{14,\mathrm{info}}-T_{23,\mathrm{info}})$
time steps. If after this waiting time a jump to its target site is
not possible (s)he will switch to the other route.

This algorithm is slightly more complicated in the 5link system but
works in the same sense: if an agent, due to a random decision,
chose a route which does not have the lowest expected travel time
and this route is blocked, (s)he will immediately decide for another
route. If the chosen route does have the lowest expected travel time
and the routes' entrance is blocked, on $j_1$ the agent will wait
$\kappa_{j_1,\mathrm{thres}}$ times the to-be-expected saved travel
time before switching. In the 5link network, an analogous algorithm
operates at junction $j_2$ where the parameter
$\kappa_{j_2,\mathrm{thres}}$ is introduced. For more details, see
\cite{bittihn2018phd} where the algorithms for decision making are
shown in pseudo code.

In the following the different types of information that are used
for the algorithm are explained. Furthermore the relaxation
procedures used in the simulations are explained.

\subsection{Public historical information}

Public historical information is implemented as follows: each time
any user finishes one round (jumps out of $j_4$), the experienced
travel time is recorded. This travel time is then made available to
all agents as their $T_{\mathrm{info}}$. This information is
historical since, as explained in Sec.~\ref{sec:types_traffic_info},
the traffic state might have changed during the round. For this type
of information a short relaxation process is needed: at the
beginning all agents follow randomly assigned routes. As soon as
each route has been used at least once, the system is considered to
be relaxed. It has already been shown that this type of information
does not lead to user optima in various two route scenarios but
rather to very strong oscillations around the user optimum
(e.g.~\cite{wahle2000decision}). In~\cite{bittihn2018phd} we show
that the expected oscillations in the 4link system / two route
scenario can be reproduced. Furthermore it is found that a
similar behaviour occurs in the 5link / three route scenario. We do
not present these results here, since they do not offer new
insights. The interested reader is referred
to~\cite{bittihn2018phd}.

\subsection{Public predictive information}
Public predictive information is provided on the basis of the
current positions of all agents in the network. It is implemented as
an approximation of the traffic information provided by smartphone
apps in real road networks. To provide estimates of travel times for
all edges, the densities $\rho_i$ are determined from the current
number of particles on each edge $E_i$: $\rho_i=n_i/L_i$ where $n_i$
is the number of particles on edge $E_i$. From this density a travel time
prediction $T_{i,\mathrm{pred}}$ is calculated employing the formula
\begin{equation}
 T_{i,\mathrm{pred}}=\frac{L_i}{1-\rho_i}. \label{eq:tt_per_tasep}
\end{equation}
This equation is the exact \textit{stationary state} expression for
the travel time of a particle in a TASEP of length $L_i$
with periodic boundary conditions and density
$\rho_i$~\cite{bittihn2016}. In our case it is only an approximation
for the travel time on an edge: the edges neither have periodic
boundary conditions nor are they (necessarily) in stationary
states. It will show to produce reasonably accurate approximations
(at least at low global densities) in the Braess network. From the
approximated travel times of all edges the expected route travel
times, which are used as the traffic information for all agents in
the route choice algorithm, are calculated as
\begin{eqnarray}
 T_{14,\mathrm{info}}&=T_{1,\mathrm{pred}}+T_{4,\mathrm{pred}}, \label{eq:t14_info_publ_pred}\\
 T_{23,\mathrm{info}}&=T_{2,\mathrm{pred}}+T_{3,\mathrm{pred}}, \label{eq:t23_info_publ_pred}\\
 T_{153,\mathrm{info}}&=T_{1,\mathrm{pred}}+T_{5,\mathrm{pred}}+T_{3,\mathrm{pred}}.
 \label{eq:t153_info_publ_pred}
\end{eqnarray}
Here further approximations are introduced: potential waiting times
at the junction sites are not considered and the system is only
described in a mean field fashion since correlations are neglected.

At each decision point the current $T_{i,\mathrm{info}}$ is
calculated based on the current positions of all particles. This
means that if an agent bases its decision on the
$T_{i,\mathrm{info}}$ before starting a round and then re-decides on
one of the junctions, the $T_{i,\mathrm{info}}$ might already have
changed at the time of the second decision.

For this type of information no relaxation process is required, the
agents are just placed on random positions in the network.

\subsection{Personal historical information}
Personal historical information is information based on the agents'
experiences from previous rounds. Each agent is assigned a memory
capacity of $c_{\mathrm{mem}}$ rounds. For the last
$c_{\mathrm{mem}}$ rounds, each agent remembers which routes it took
and their corresponding travel times. From these times the
$T_{i,\mathrm{info}}$ are calculated: they are the mean values of
the travel times of each route as experienced in the last
$c_{\mathrm{mem}}$ rounds. Additionally, each agent remembers its
last experienced travel times on all three routes. Like this, even
if e.g. route 23 was not used in the last $c_{\mathrm{mem}}$ rounds,
the agent will still remember the travel time of that route from the
last usage (that lies more than $c_{\mathrm{mem}}$ rounds in the
past).

For this kind of information there is a two-fold relaxation process.
First, each agent is placed on a random position with a random
strategy. Then it tries to gather one travel time value for each
route. Once these values are obtained, the system keeps evolving
until each agent has experienced $c_{\mathrm{mem}}$ rounds. Once
each agent used each route at least once and has a filled its memory
of capacity of $c_{\mathrm{mem}}$ rounds, the system is considered
relaxed.


\section{Simulation results}
\label{sec:results}

We applied the algorithms described in the previous section to
determine the effects of the route choice behavior in the Braess
network with TASEP-based traffic dynamics. The edge lengths
$L_0,\ldots,L_4$ are given by Eq.~(\ref{eq:edgelenghts}). We
focussed on four different combinations of $L_5$, the length of the
added edge, and $M$, the number of agents. These are indicated by
$\star 1$ to $\star 4$ in Figs.~\ref{fig:phase_diagrams}~(a)
and~(b). We applied the route choice algorithm to these states to
see if user optimum states attainable by externally tuning all
agents' strategies are realized. $\star 1$ and $\star 2$ have been
chosen as representatives of states in which the new road is
expected to improve the traffic situation: For both types of
externally tuned strategies, i.e.\ for fixed numbers of agents
following fixed routes and for all cars deciding according to the
same turning probabilities, in states $\star ~1$ and $\star2$ the
new road leads to a 5link system with lower user optimum travel
times than in the corresponding 4link system. State $\star1$ is a
special state since in the 5link user optimum all cars use new route
while the old routes have higher travel times and are not used. In
the user optima of state $\star 2$ all routes are used. State
$\star3$ is a Braess state for externally tuned strategies: the user
optima in the 5link networks have higher travel times than those of
the corresponding 4link networks' user optima. For fixed personal
strategies, in state $\star3$ two user optima exist. State $\star4$
is a Braess state for externally tuned strategies and fixed personal
strategies. For externally tuned turning probabilities no (short
term) user optima exist in this state since fluctuating domain walls
are found in the system.

More details about the described states of the four points $\star 1$
to $\star 4$ are found in~\cite{bittihn2016,bittihn2018}. Details
for the exact parameters of the four states, their user optima and
the corresponding travel times of these user optima can be found
in~\ref{sec:app_test_states}. These details are not essential
if one is interested in qualitative effects, but might be helpful
for a quantitative understanding of the following results.

For the obtained results which are discussed in the remainder of the
present section, the parameters of the algorithm as introduced in
Section~\ref{sec:route_choice_alg} were always chosen to be:
\begin{eqnarray}
 p_{\mathrm{info}}&=0.9, \\
 \Delta T_{\mathrm{thres}}&=10, \\
 \kappa_{j_1,\mathrm{thres}}=\kappa_{j_2,\mathrm{thres}}&=0.1, \\
 c_{\mathrm{mem}}&=30~,
\end{eqnarray}
where $c_{\mathrm{mem}}$ is only needed for the case of
personal historic information. As we consider the continuous route
choices in the Braess network to mirror a commuter scenario, it is
reasonable to assume that in the majority of times, drivers choose
their routes with the aim of minimizing their travel times, i.e.
$p_{\mathrm{info}}=0.9$. Furthermore it appears to be realistic that
a commuter remembers approximately one month of her last experienced
travel times ($c_{\mathrm{mem.}}=30$). Assuming that drivers
may not switch from their preferred route if the expected saved time
is really low, a value of $\Delta T_{\mathrm{thres}}=10$ seems
reasonable.

Figs.~\ref{fig:publ_pred_star1}--\ref{fig:pers_hist_star4} show the
results for all four states $\star 1$ to $\star 4$, each with public
predictive and personal historical information. All figures have
parts~(a) and~(b). Parts~(a) show how the mean values of the travel
times $\bar{T}_i$ of the routes $i$ develop with the system time
where all times are measured in numbers of performed Monte Carlo
sweeps. For details on the simulation process the reader is referred
to~\cite{bittihn2018phd}. The values of both routes in the 4link
system and of the three routes in the corresponding 5link systems
are shown for comparison. One can thus see if the new road leads to
higher or lower travel times. Additionally the travel times that are
expected from the pure and mixed user optima of the 4link and 5link
systems with externally tuned strategies are shown for comparison by
the dotted grey lines whose values are given by the $\tau_i$ on the
second $y$-axis on the right.

Parts~(b) of
Figs.~\ref{fig:publ_pred_star1}--\ref{fig:pers_hist_star4} show the
two variables $m_{\mathrm{l}}^{(j_1)}=1-M_{23}/M$ and
$m_{\mathrm{l}}^{(j_2)}=M_{14}/(M_{14}+M_{153})$~\footnote{In the
  4link system only $m_{\mathrm{l}}^{(j_1)}$ is needed which reduces
  to $m_{\mathrm{l}}^{(j_1)}=M_{14}/M$} against the system
time\footnote{In~\cite{bittihn2018phd} these two
    variables are called implicit turning probabilities
    $\gamma_{\mathrm{imp.}}$ and $\delta_{\mathrm{imp.}}$.}.  Here,
the $M_{14},~M_{23}$ and $M_{153}$ are the numbers of cars which
follow routes 14, 23 and 153 at that system time. The values of the
two variables $m_{\mathrm{l}}^{(j_1)}$ and $m_{\mathrm{l}}^{(j_2)}$
capture the fractions of agents using the
three different routes. Similar to the variables
$n_{\mathrm{l}}^{(j_1)}$ and $n_{\mathrm{l}}^{(j_2)}$, which we
introduced in Eqs.~(\ref{eq:nlj1}),(\ref{eq:nlj2}) in the context of
externally tuned fixed personal strategies, they describe the
fraction of particles turning `left' of junctions $j_1$ and $j_2$,
\textit{but in the present moment}. They represent the strategies
that the particles choose as a result of the route choice algorithm.
Due to the algorithm their values can change before and during the
rounds.

The $m_{\mathrm{l}}^{(j_1)}$ and $m_{\mathrm{l}}^{(j_2)}$ can be
compared to the $(n_{\mathrm{l}}^{(j_1)},n_{\mathrm{l}}^{(j_2)})$
that realize the pure user optima for externally tuned fixed
personal strategies and the $(\gamma,\delta)$ that realize the mixed
user optima for externally tuned turning probabilities. Those
strategies realizing the user optima by externally tuning the route
choices are also shown by the dotted grey lines whose values are
given by the $\sigma_i$ on the second $y$-axis on the right of
parts~(b) of
Figs.~\ref{fig:publ_pred_star1}--\ref{fig:pers_hist_star4}. This allows
to determine whether the algorithm drives the system
into a user optimum. This gives an indication on how close it is to
an expected pure or mixed optimum. To see if real pure user optima
or real mixed user optima are realized, further analysis is needed:
in a real pure user optimum no individual users would switch routes
any more (this can not be seen from the $m_{\mathrm{l}}^{(j_1)}$ and
$m_{\mathrm{l}}^{(j_2)}$, since they only show the sums of particles
following specific routes). Situations without any individual
particles switching routes can not be obtained in our algorithm
since $p_{\mathrm{info}}<1$. Still, if the number of switches is low
one can presume that the algorithm brings the system close to a pure
user optimum. In a mixed user optimum one would expect a higher
number of individual route switches. To test if indeed a real mixed
user optimum is realized a further statistical analysis of the
behaviour of all individual particles is needed. Such an analysis is
not shown here. The interested reader is referred
to~\cite{bittihn2018phd}, where it is performed (in parts) for the
state $\star 3$.

The results presented in the following all show single Monte Carlo
simulations of the systems. They were confirmed in numerous runs of
the same instances of the system with different RNG seeds, which
showed the same behaviour apart from minor variations due to the
stochasticity of the process.

\subsection{Public predictive information.}

\begin{figure}[ht]
  \centering
  \includegraphics{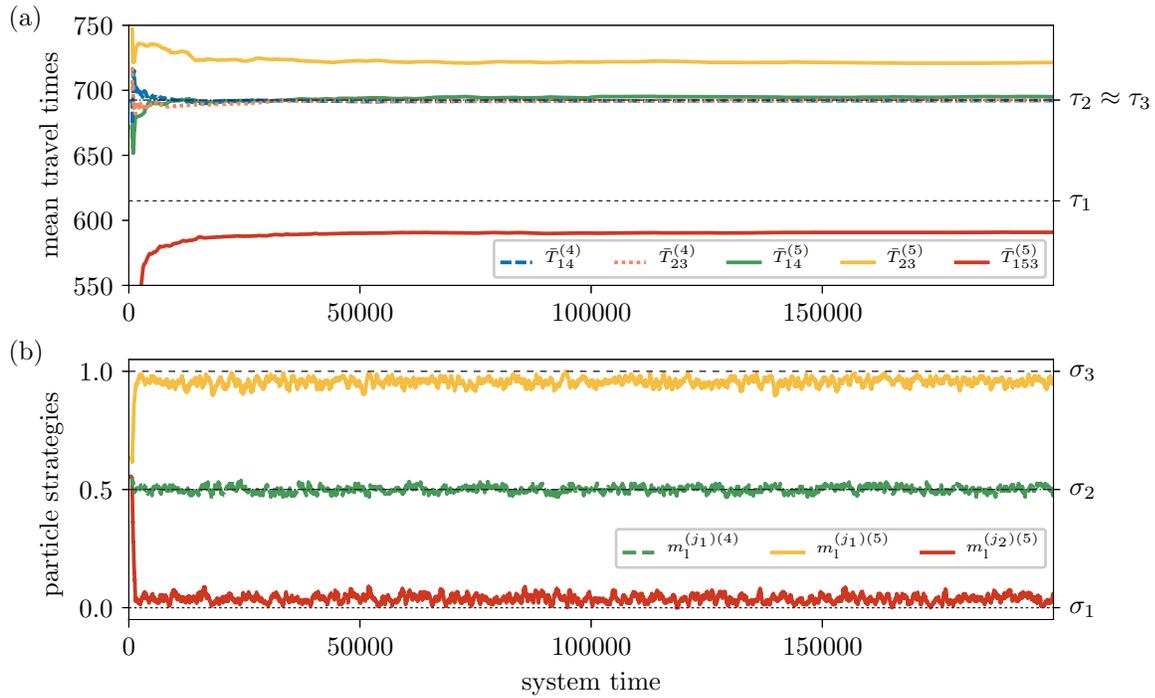}
  \caption{\label{fig:publ_pred_star1}Results in state $\star1$
    if public predictive information is provided. As seen in (a), the
    mean travel times of both routes in the 4link system coincide with
    those from the expectations. The travel time of
    route 153 in the 5link is the lowest of the three routes. The expected travel times
    in the pure and mixed user optima found by externally tuning the strategies are shown by
    the dotted lines with $T^{(4)}_{\mathrm{max,puo}}=\tau_2$, $T^{(4)}_{\mathrm{max,muo}}=\tau_3$,
    $T^{(5)}_{\mathrm{max,puo}}=T^{(5)}_{\mathrm{max,muo}}=\tau_1$.
    Part (b) shows that in the 4link both routes are, as expected, used by
    equal amounts of agents, and in the 5link almost all cars use
    route 153. The strategies realizing the user
    optima by externally tuning the strategies are given for
    comparison with $n_{\rm l,puo}^{(j_1)(4)}=\gamma_{\rm muo}^{(4)}=\sigma_2$,
    $\left(n_{\rm l,puo}^{(j_1)(5)},n_{\rm l,puo}^{(j_2)(5)} \right)=
    \left(\gamma_{\rm muo}^{(5)},\delta_{\rm muo}^{(5)} \right)=\left(\sigma_3,\sigma_1\right)$.
    The ``$E_5$ optimal, all 153" state is realized.
        }
\end{figure}

\begin{figure}[ht]
  \centering
  \includegraphics{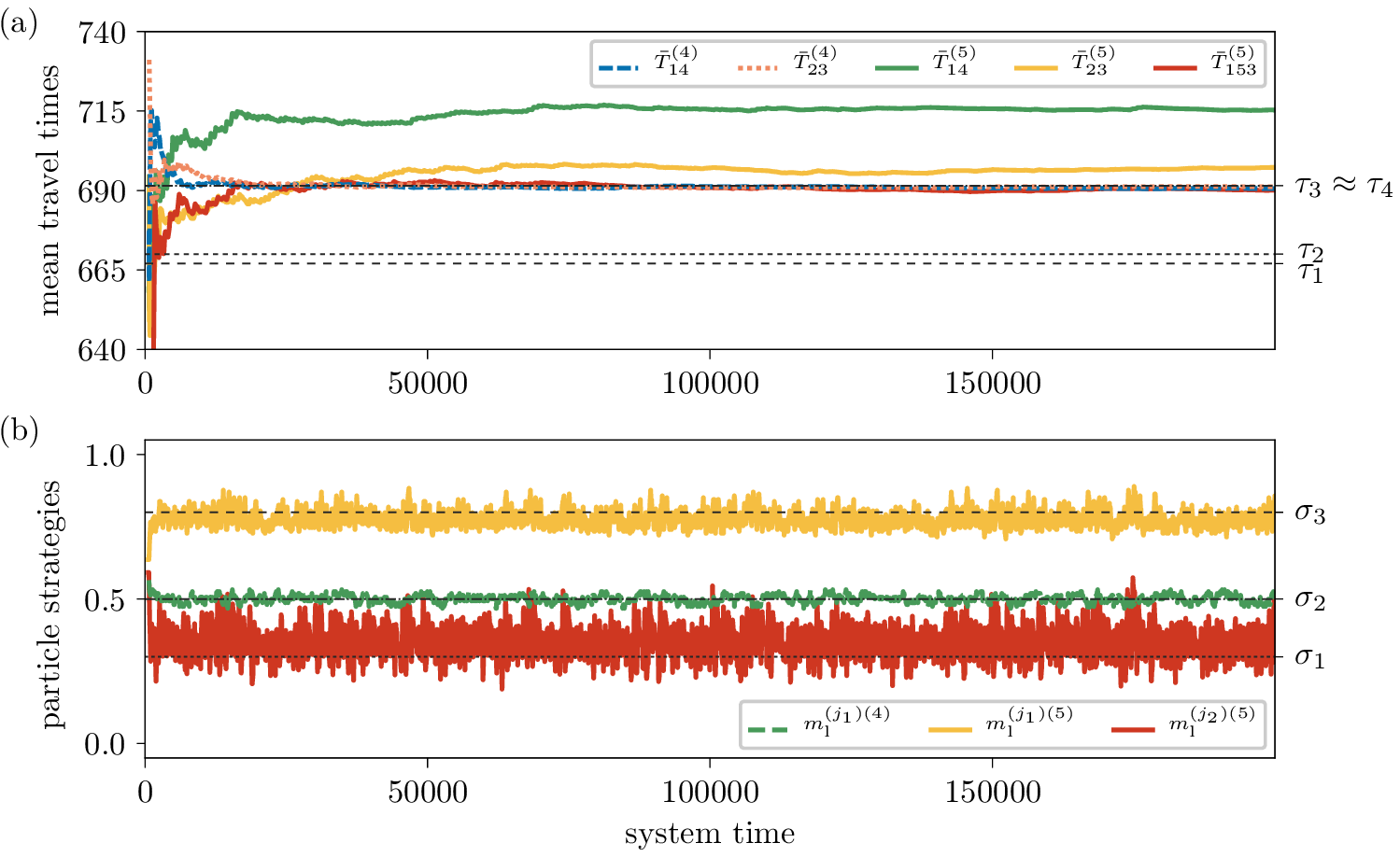}
  \caption{\label{fig:publ_pred_star2}Results for state $\star2$
    if public predictive information is provided. As seen in (a), the
    mean travel times of both routes in the 4link system coincide with
    those from the expectations. In the 5link system, all three routes
    have higher mean travel times than expected in the pure and mixed
    5link user optima. The travel times are also higher than those in
    the 4link. The expected travel times in the pure and mixed user optima found by
    externally tuning the strategies are shown by the dotted lines with
    $T^{(4)}_{\mathrm{max,puo}}=\tau_3$, $T^{(4)}_{\mathrm{max,muo}}=\tau_4$,
    $T^{(5)}_{\mathrm{max,puo}}=\tau_2$,
    $T^{(5)}_{\mathrm{max,muo}}=\tau_1$.
    Part (b) shows that in the 4link both routes are, as
    expected, used by equal amounts of agents. In the 5link system
    fluctuations around the expected user optima are observed.
    The strategies realizing the user optima by externally tuning the strategies are given for
    comparison with $n_{\rm l,puo}^{(j_1)(4)}=\gamma_{\rm muo}^{(4)}=\sigma_2$,
    $\left(n_{\rm l,puo}^{(j_1)(5)},n_{\rm l,puo}^{(j_2)(5)} \right)=
    \left(\gamma_{\rm muo}^{(5)},\delta_{\rm muo}^{(5)} \right)=\left(\sigma_3,\sigma_1\right)$.
    Judging from the mean travel time values, a Braess state instead of the
    expected ``$E_5$ optimal" state is observed.
    }
\end{figure}

\begin{figure}[ht]
  \centering
  \includegraphics{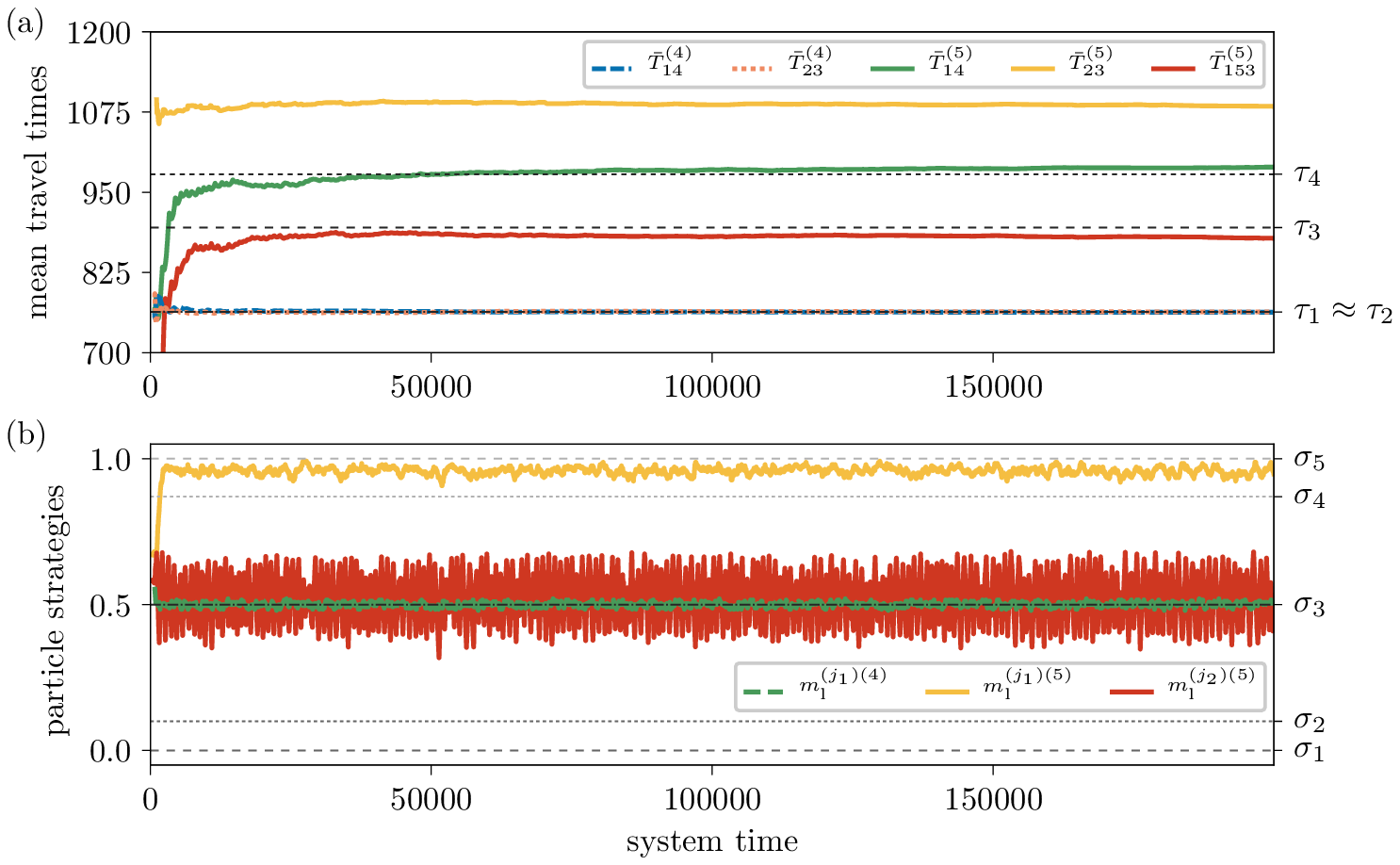}
  \caption{\label{fig:publ_pred_star3}Results for state $\star3$
    if public predictive information is provided. As seen in (a), the
    mean travel times of both routes in the 4link system coincide with
    those from the expectations. In the 5link system, all three routes
    have unequal travel times, all higher than those in the
    4link. Travel times of route 14 and 153 are similar and close to
    those expected in the 5link user optima. The expected travel times
    in the pure and mixed user optima found by externally tuning the strategies
    are shown by the dotted lines with $T^{(4)}_{\mathrm{max,puo}}=
    \tau_2$, $T^{(4)}_{\mathrm{max,muo}}=\tau_1$,
    $T^{(5)}_{\mathrm{max,puo(i)}}=T^{(5)}_{\mathrm{max,puo(ii)}}=\tau_4$,
    $T^{(5)}_{\mathrm{max,muo}}=\tau_3$.
    Part (b) shows that in the 4link both routes are, as expected, used by equal amounts of
    particles. In the 5link system strong fluctuations around one of
    the three theoretically accessible user optima, i.e. around the
    pure user optimum $\mathrm{puo(ii)}$, are observed. Route 23 is
    not used by many particles. The strategies realizing the user
    optima by externally tuning the strategies are given for
    comparison with $n_{\rm l,puo}^{(j_1)(4)}=\gamma_{\rm muo}^{(4)}=\sigma_3$,
    $\left(n_{\rm l,puo(i)}^{(j_1)(5)},n_{\rm l,puo(i)}^{(j_2)(5)} \right)=\left(\sigma_3,\sigma_1\right)$,
    $\left(n_{\rm l,puo(ii)}^{(j_1)(5)},n_{\rm l,puo(ii)}^{(j_2)(5)} \right)=\left(\sigma_5,\sigma_3\right)$,
    $\left(\gamma_{\rm muo}^{(5)},\delta_{\rm muo}^{(4)} \right)=\left(\sigma_4,\sigma_2\right)$.
    The ``Braess 1" state is realized on average.
    }
\end{figure}

\begin{figure}[ht]
  \centering
  \includegraphics{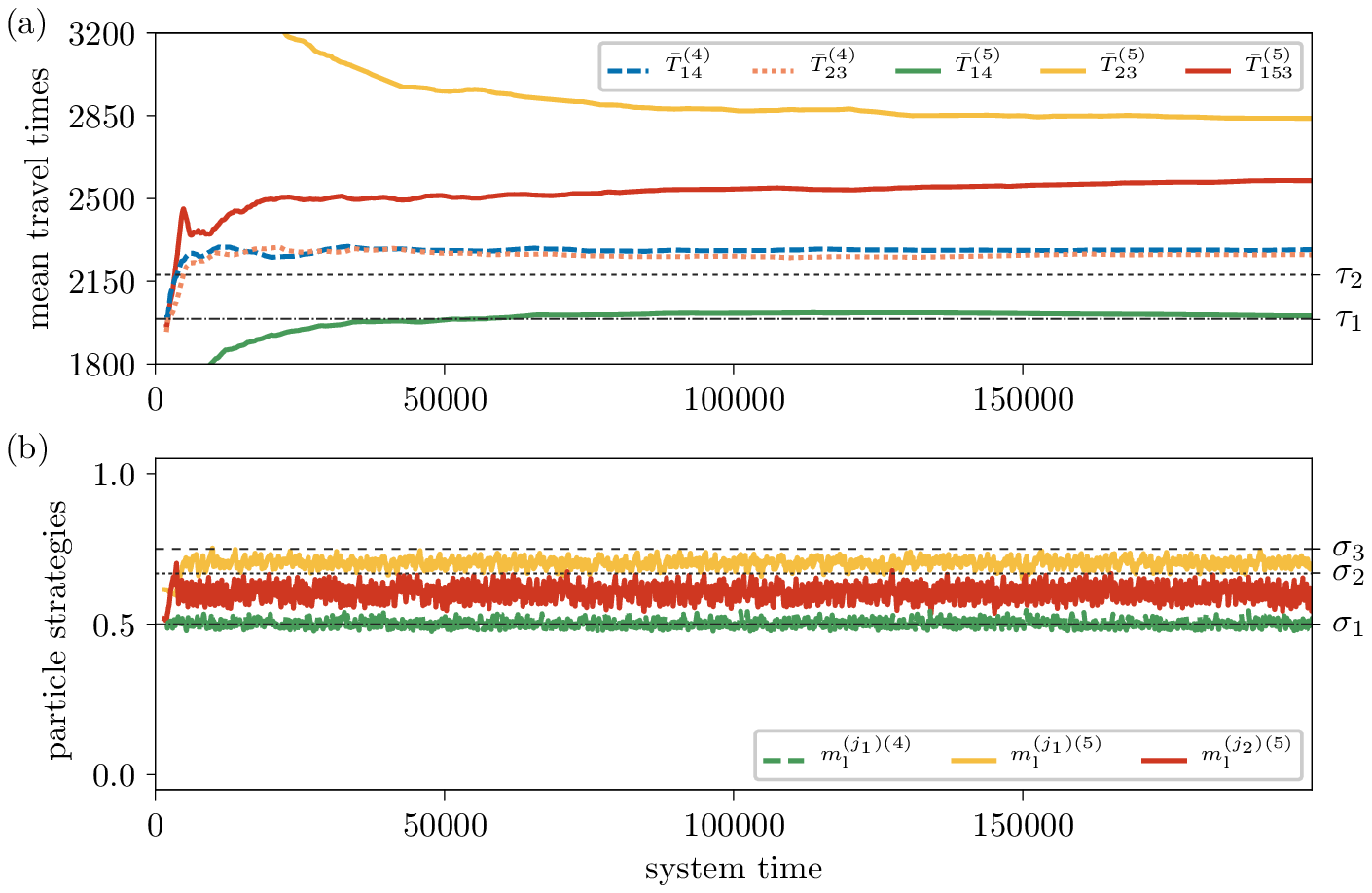}
  \caption{\label{fig:publ_pred_star4}Results for state $\star4$
    if public predictive information is provided. As seen in (a), the
    mean travel times of both routes in the 4link system equalize, but
    at a slightly higher value than the expectations. In the 5link
    system, all three routes have unequal travel times, route 23 and
    153 higher than those in the 4link, route 14 lower. The expected travel times
    in the pure and mixed user optima found by externally tuning the strategies are
    shown by the dotted lines with $T^{(4)}_{\mathrm{max,puo}}=\tau_1$,
    $T^{(5)}_{\mathrm{max,puo}}=\tau_2$.
    Part (b) shows that in the 4link both routes are, as expected, used by equal
    amounts of agents. In the 5link system strong fluctuations
    around the theoretically accessible pure user optima are
    observed. The strategies realizing the user optima by externally tuning the
    strategies are given for comparison with $n_{\rm l,puo}^{(j_1)(4)}=\sigma_1$,
    $\left(n_{\rm l,puo}^{(j_1)(5)},n_{\rm l,puo}^{(j_2)(5)} \right)=\left(\sigma_3,\sigma_2\right)$.
    }
\end{figure}

The results for the algorithm with public predictive information are
shown in Figs.~\ref{fig:publ_pred_star1}--\ref{fig:publ_pred_star4}
for the four different states $\star1$ to $\star4$.
Fig.~\ref{fig:publ_pred_star1} shows results for state $\star 1$.
One can see that the user optima of both the 4link and the 5link
networks are realized in a stable manner.
In the 4link network approximately half the agents choose route 14
and the other half route 23 (Fig.~\ref{fig:publ_pred_star1}~(b)).
Their mean travel times (Fig.~\ref{fig:publ_pred_star1}~(a))
equalize at the value expected from the user optima obtained in
networks with externally tuned strategies. In the 5link network,
apart from small fluctuations, almost all agents choose route 153
(Fig.~\ref{fig:publ_pred_star1}~(b)) and this route has a lower
travel time than the other two (almost unused) routes in the 5link
and also lower than those in the 4link system
(Fig.~\ref{fig:publ_pred_star1}~(a)). The ``$E_5$ optimal, all 153"
state that is expected is thus realized.

The results for state $\star 2$ are shown in
Fig.~\ref{fig:publ_pred_star2}. In the 4link system a user optimum
is realized. We can see that approximately $M/2$ agents use routes
14 and 23 without large fluctuations
(Fig.~\ref{fig:publ_pred_star2}~(b)). The mean travel times of the
two routes in the 4link system thus equalize
(Fig.~\ref{fig:publ_pred_star2}~(a)).
In the 5link system a different behaviour is observed. In
Fig.~\ref{fig:publ_pred_star2}~(b) it can be seen that the expected
user optimum is reached \textit{on average}. The numbers of cars on
the three routes (as represented through the
$m_{\mathrm{l}}^{(j_1)(5)}$ and $m_{\mathrm{l}}^{(j_2)(5)}$ in
Fig.~\ref{fig:publ_pred_star2}~(b)) oscillate around the values
expected from the
$(n_{\mathrm{l,puo}}^{(j_1)(5)},n_{\mathrm{l,puo}}^{(j_2)(5)})$ and
$(\gamma^{(5)}_{\mathrm{muo}},\delta^{(5)}_{\mathrm{muo}})$ from the
pure and mixed user optima for externally tuned strategies. Due to
these fluctuations the mean travel times of the three routes are
close to each other but not equal
(Fig.~\ref{fig:publ_pred_star2}~(a)). Opposed to the expectation,
the mean travel times of routes 14 and 23 are actually higher than
those in the 4link system. Thus, even if the user optimum of the
5link system is realized on average, the algorithm with public
predictive information drives the system into a state which is more
of `Braess nature' in the sense that the 5link travel times are
higher than the 4link's, opposed to the expected ``$E_5$ optimal"
state.

Fig.~\ref{fig:publ_pred_star3} shows results for the case of public
predictive information in state $\star 3$.
As in the two previous states, the user optimum of the 4link system
is realized. As seen in Figure~\ref{fig:publ_pred_star3}~(b) half of
the particles use routes 14 and 23. As can be seen in
Figure~\ref{fig:publ_pred_star3}~(a) the travel times of both routes
equalize at the expected value. As detailed
in~\ref{sec:app_test_states}, the 5link system of state $\star 3$
with externally tuned parameters has two pure user optima and one
mixed user optimum. The values of the $(n_{\mathrm{l,
puo(i/ii)}}^{(j_1)(5)},n_{\mathrm{l, puo(i/ii)}}^{(j_2)(5)})$ of
neither of the two pure optima puo(i) and puo(ii) coincides with the
values of the $(\gamma_{\rm muo}^{(5)}, \delta_{\rm muo}^{(5)})$ of
the mixed user optimum muo. In the 5link system the route choice
algorithm produces strong fluctuations around the pure user optimum
$\mathrm{puo(ii)}$ (Fig.~\ref{fig:publ_pred_star3}~(b)), a state in
which route 23 is not used and half the agents choose route 14 and
the other half route 153. Due to the fluctuations around the user
optimum the travel times of the two used routes (route 14 and 153)
are close to each other but not exactly equal
(Fig.~\ref{fig:publ_pred_star3}~(a)). They are all higher than those
of the two routes in the 4link system. Thus a Braess state is
realized. The almost unused route 23 has an even higher travel time,
as expected.

Fig.~\ref{fig:publ_pred_star4} shows results for state $\star 4$. In
the 4link system the user optimum is realized. Even if fluctuations
around the 4link user optimum are also small in state $\star 4$, the
travel times of both routes equalize at a slightly higher value than
expected (Fig.~\ref{fig:publ_pred_star4}~(a)). This is a consequence
of jamming effects in front of $j_4$ which play a larger role at
higher densities.
In the 5link network with externally tuned particles only a pure
user optimum exists (see~\ref{sec:app_test_states}). In the system
with externally tuned turning probabilities fluctuating domain walls
are observed at such high densities and thus no (short term) user
optimum exists~\cite{bittihn2016}. The route choice algorithm drives
the system close to the pure user optimum. The resulting
$m_{\mathrm{l}}^{(j_1)(5)}$ and $m_{\mathrm{l}}^{(j_2)(5)}$ are
slightly different from the expected $(n_{\mathrm{l,
puo}}^{(j_1)(5)},n_{\mathrm{l, puo}}^{(j_2)(5)})$ in the pure user
optimum. Furthermore they fluctuate
(Fig.~\ref{fig:publ_pred_star4}~(b)). Thus the travel times of the
three routes are not equal. Routes 23 and 153 have higher travel
times than the routes of the 4link system
(Fig.~\ref{fig:publ_pred_star4}~(b)), which can be interpreted as a
kind of Braess behaviour.

One can conclude that the route choice based on public predictive
information typically drives the system into user optima in the
4link systems. This is not surprising if one remembers how the
predicted travel times are calculated (see
Eqs.~(\ref{eq:t14_info_publ_pred})--(\ref{eq:t153_info_publ_pred})):
since the 4link system is symmetric and the pure user optima are
always given for an equal distribution onto both routes, the
algorithm which counts the numbers of particles for its travel time
predictions will always realize such user optima.

In the 5link networks user optima are not always realized since road
5 breaks the symmetry of the 4link network. In the 5link network,
the public predictive information realizes user optima at low global
densities. This is the case since in this density regime the
correlations between the roads do not influence the route travel
times strongly. At higher densities the correlations become more
important leading to traffic jams near junction sites. Here the
travel time predictions become less accurate, leading to
fluctuations around the user optima. This can be seen
in~\ref{sec:how_well_publ_pred_info}, where the accuracy of the
predicted travel times is shown.

\subsection{Personal historical information.}

\begin{figure}[ht]
  \centering
  \includegraphics{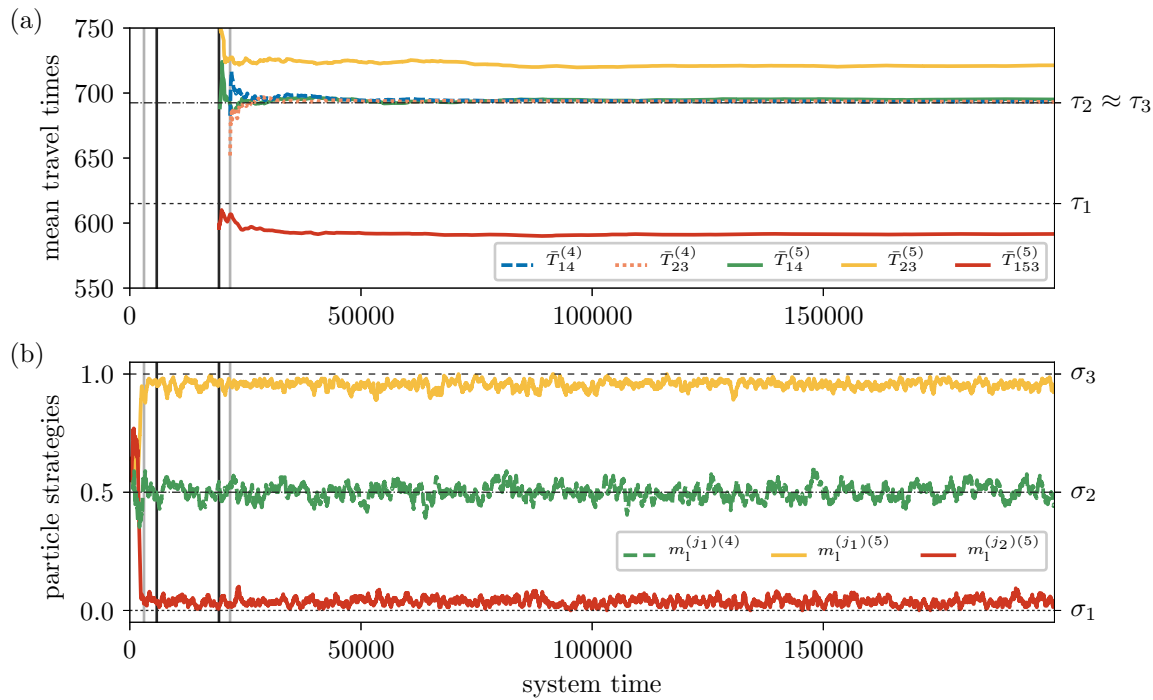}
  \caption{\label{fig:pers_hist_star1}Results for state $\star1$
    if personal historical information is provided. As seen in (a),
    the mean travel times on the routes in the 4link system equalize
    at the expected values. In the 5link system, the travel time on
    route 153 is slightly lower than expected. The expected travel times
    in the pure and mixed user optima found by externally tuning the strategies
    are shown by the dotted lines with $T^{(4)}_{\mathrm{max,puo}}=
    \tau_2$, $T^{(4)}_{\mathrm{max,muo}}=\tau_3$,
    $T^{(5)}_{\mathrm{max,puo}}=T^{(5)}_{\mathrm{max,muo}}=\tau_1$.
    Part (b) shows that in the 4link and 5link networks, the strategies develop as expected
    in the theoretically accessible user optima. In the 5link, route
    153 is used almost exclusively. The strategies realizing the user
    optima by externally tuning the strategies are given for comparison with
    $n_{\rm l,puo}^{(j_1)(4)}=\gamma_{\rm muo}^{(4)}=\sigma_2$,
    $\left(n_{\rm l,puo}^{(j_1)(5)},n_{\rm l,puo}^{(j_2)(5)} \right)=
    \left(\gamma_{\rm muo}^{(5)},\delta_{\rm muo}^{(5)} \right)=\left(\sigma_3,\sigma_1\right)$.
    The ``$E_5$ optimal, all 153" state is realized.
    }
\end{figure}

\begin{figure}[ht]
  \centering
  \includegraphics{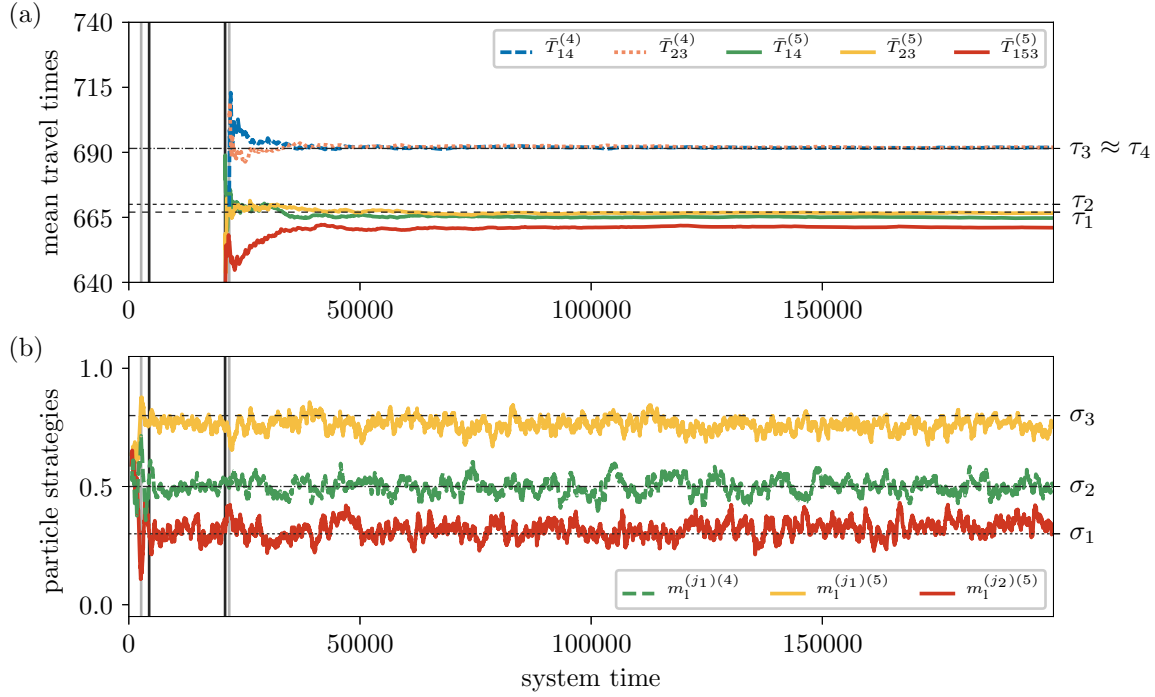}
  \caption{\label{fig:pers_hist_star2}Results for state $\star2$
    if personal historical information is provided. As seen in (a),
    the mean travel times on the routes in the 4link and 5link systems
    equalize at the expected values. The expected travel times
    in the pure and mixed user optima found by externally tuning the strategies are
    shown by the dotted lines with $T^{(4)}_{\mathrm{max,puo}}=\tau_3$,
    $T^{(4)}_{\mathrm{max,muo}}=\tau_4$,
    $T^{(5)}_{\mathrm{max,puo}}=\tau_2$,
    $T^{(5)}_{\mathrm{max,muo}}=\tau_1$.
    Part (b) shows that in the 4link
    and 5link networks, the strategies develop as expected in the
    theoretically accessible user optima. The strategies realizing the user
    optima by externally tuning the strategies are given for
    comparison with $n_{\rm l,puo}^{(j_1)(4)}=\gamma_{\rm muo}^{(4)}=\sigma_2$,
    $\left(n_{\rm l,puo}^{(j_1)(5)},n_{\rm l,puo}^{(j_2)(5)} \right)=
    \left(\gamma_{\rm muo}^{(5)},\delta_{\rm muo}^{(5)} \right)=\left(\sigma_3,\sigma_1\right)$.
    The ``$E_5$ optimal" state is realized.}
\end{figure}

\begin{figure}[ht]
  \centering
  \includegraphics{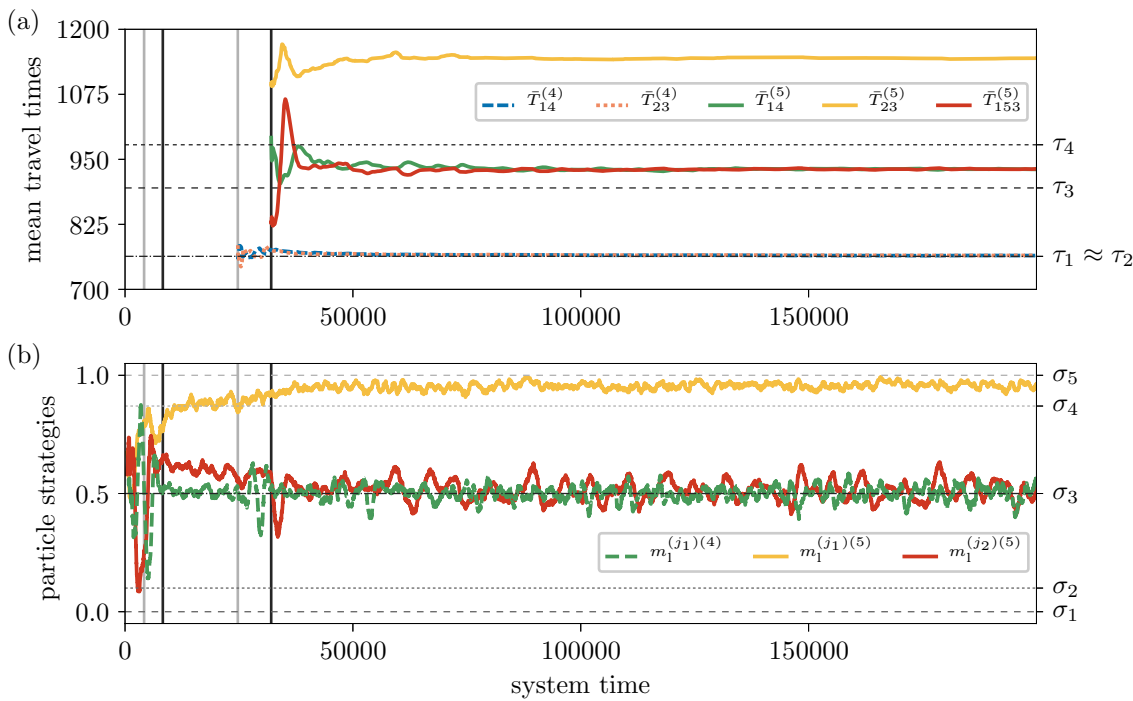}
  \caption{\label{fig:pers_hist_star3}Results for state $\star3$
    if personal historical information is provided. As seen in (a),
    the mean travel times of the routes in the 4link system equalize
    at the expected values. In the 5link system, the mean travel times
    of routes 14 and 153 equalize at a value lower than that of route
    23, but higher than those of the 4link system. The expected travel times
    in the pure and mixed user optima found by externally tuning the strategies
    are shown by the dotted lines with
    $T^{(4)}_{\mathrm{max,puo}}=\tau_2$, $T^{(4)}_{\mathrm{max,muo}}=\tau_1$,
    $T^{(5)}_{\mathrm{max,puo(i)}}=T^{(5)}_{\mathrm{max,puo(ii)}}=
    \tau_4$, $T^{(5)}_{\mathrm{max,muo}}=\tau_3$.
    Part (b) shows that in the 4link the strategies develop as expected in the
    theoretically accessible user optima. In the 5link system, the
    pure user optimum $\mathrm{puo(ii)}$ is approached with some small
    fluctuations. In the 5link system, routes 14 and 153 are used by
    approximately half the agents each, while route 23 is almost not
    used at all. The strategies realizing the user
    optima by externally tuning the strategies are given for
    comparison with $n_{\rm l,puo}^{(j_1)(4)}=\gamma_{\rm muo}^{(4)}=\sigma_3$,
    $\left(n_{\rm l,puo(i)}^{(j_1)(5)},n_{\rm l,puo(i)}^{(j_2)(5)} \right)=\left(\sigma_3,\sigma_1\right)$,
    $\left(n_{\rm l,puo(ii)}^{(j_1)(5)},n_{\rm l,puo(ii)}^{(j_2)(5)} \right)=\left(\sigma_5,\sigma_3\right)$,
    $\left(\gamma_{\rm muo}^{(5)},\delta_{\rm muo}^{(4)} \right)=\left(\sigma_4,\sigma_2\right)$.
    The ``Braess 1" state is realized.
    }
\end{figure}

\begin{figure}[ht]
  \centering
  \includegraphics{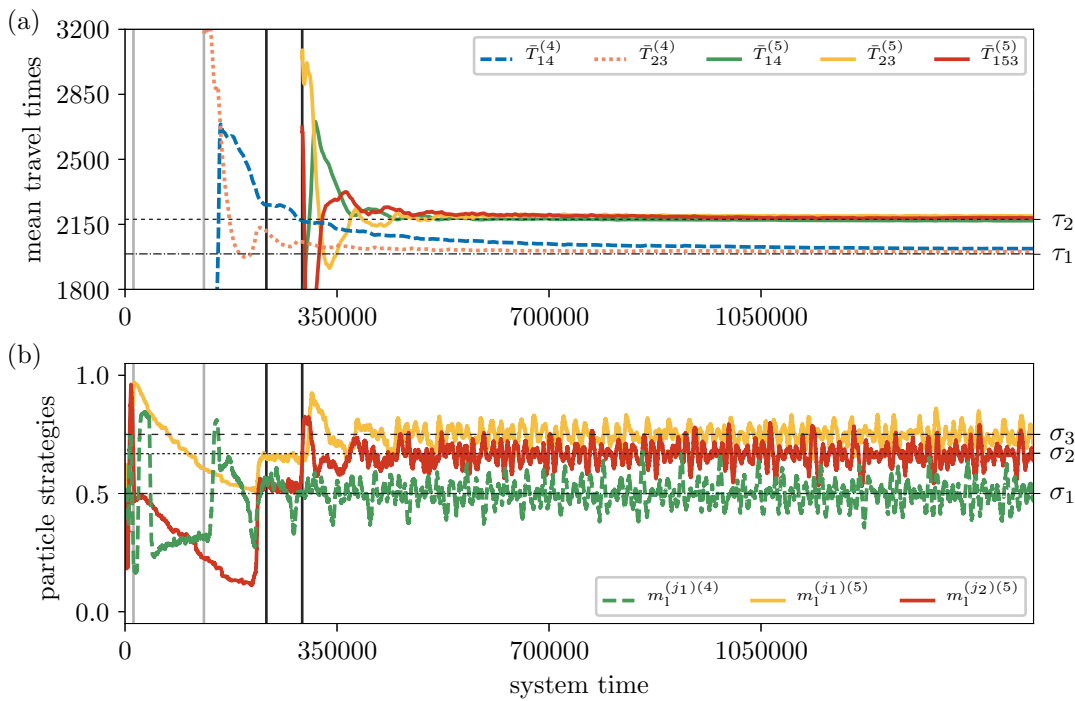}
  \caption{\label{fig:pers_hist_star4}Results for state $\star4$
    if personal historical information is provided. In this state with
    higher global density than states $\star1$ to $\star3$ a longer
    relaxation time is needed, which is why a longer system time
    period is plotted as compared to Figs.~\ref{fig:publ_pred_star1}
    to~\ref{fig:pers_hist_star3}. As seen in (a), the mean travel
    times of the routes in the 4link and 5link systems equalize at the
    expected values. The expected travel times
    in the pure and mixed user optima found by externally tuning the
    strategies are shown by the dotted lines with $T^{(4)}_{\mathrm{max,puo}}=\tau_1$,
    $T^{(5)}_{\mathrm{max,puo}}=\tau_2$.
    Part (b) shows that in the 4link and 5link the strategies develop as expected
    in the theoretically accessible user optima.
    The strategies realizing the user
    optima by externally tuning the strategies are given for
    comparison with $n_{\rm l,puo}^{(j_1)(4)}=\sigma_1$,
    $\left(n_{\rm l,puo}^{(j_1)(5)},n_{\rm l,puo}^{(j_2)(5)} \right)
    =\left(\sigma_3,\sigma_2\right)$.
    The ``Braess 1" state, which for externally tuned
    route choices only exists with fixed route choices, is realized.
    }
\end{figure}

Results for the case of personal historical information are shown in
Figs.~\ref{fig:pers_hist_star1}--\ref{fig:pers_hist_star4} for the
four different states $\star1$ to $\star4$. In all parts~(a) and~(b)
of these figures four vertical lines are shown. The two lines in
brighter and darker grey correspond to the two relaxation times of
the 4link and 5link systems, respectively. The line further to the
left indicates the system time at which all agents have gathered at
least one travel time experience for each route. The line further to
the right indicates the system time at which the memory
capacities of all agents is full. The recording of the evolution of
the mean travel time starts once the relaxation process is finished.

Fig.~\ref{fig:pers_hist_star1} shows results for state $\star 1$.
The user optimum in the 4link system is reached with small remaining
fluctuations around the user optimum
(Fig.~\ref{fig:pers_hist_star1}~(b)). The travel times of both
routes in the 4link equalize at the expected values
(Fig.~\ref{fig:pers_hist_star1}~(a)).
In the 5link system the user optimum is also realized with some
remaining fluctuations. Almost all agents use route 153
(Fig.~\ref{fig:pers_hist_star1}~(b)) which has a lower travel time
than the other two (almost unused) routes and also a lower travel
time than the routes in the 4link system.

The results for state $\star 2$ are shown in
Fig.~\ref{fig:pers_hist_star2}. Both in the 4link and
5link system the user optimum is reached with some minor
fluctuations (Fig.~\ref{fig:pers_hist_star2}~(b)), as in
state $\star 1$.
The mean travel times of all three routes in the 5link system are
almost equal to the expected values and lower than
those of the 4link system (Fig.~\ref{fig:pers_hist_star2}~(a)). In
contrast to the algorithm with public predictive information where
the 5link travel times were higher than the 4link's and the system
thus showed Braess behaviour (Fig.~\ref{fig:publ_pred_star3}~(b)),
here the expected ``$E_5$ optimal" state is realized.

Fig.~\ref{fig:pers_hist_star3} shows results for state $\star 3$. As
in the previous states $\star 1$ and $\star 2$ the 4link user
optimum is reached with some minor fluctuations.
In the 5link system, the pure user optimum $\mathrm{puo(ii)}$
(see~\ref{sec:app_test_states}) is realized by the algorithm. Apart
from some minor fluctuations, approximately $M/2$ agents use routes
14 and 153, respectively. The travel times of these routes equalize
at a travel time below that of the almost unused route 23 and above
the travel times in the 4link system. The 5link user optimum is
reached in a more stable manner than in the system with public
predictive information (see Fig.~\ref{fig:publ_pred_star3}).

Fig.~\ref{fig:pers_hist_star4} shows results for the case of
personal historical information in state $\star 4$. One can see that
in this case a long relaxation process is needed. This is due to the
high global density: as all agents want to gather travel time
experiences for all three routes in the beginning, this leads to
routes getting blocked. The blockages are not permanent gridlocks
since agents will re-decide their route choices if they have to wait
very long at junctions $j_1$ or $j_2$. Nevertheless it takes quite
long until the whole system is relaxed. Once it is relaxed it
stabilises quickly. As can be seen in
Fig.~\ref{fig:pers_hist_star4}~(b), in the 4link system the user
optimum is reached with larger fluctuations around the expected
state than for states $\star 1$ to $\star 3$. The effect of these
larger fluctuations can also be seen in the mean travel times of the
two routes in the 4link: Fig.~\ref{fig:pers_hist_star4}~(a) shows
that the travel times' mean values only equalize after a relatively
long time.
The 5link systems' user optimum is also realized on average.
Fluctuations around the pure user optimum persist
(Fig.~\ref{fig:pers_hist_star4}~(b)). The mean travel times of the
three routes are almost equal to the expected value of the pure user
optimum, above the travel times of the 4link system. Thus, here also
a Braess state is realized.

One can conclude that user optima of both the 4link and the 5link
systems are realized in all four states by the algorithm with
personal historical information.


\section{Conclusions}

For drivers using public predictive information, user optima are
realized at low global densities. At higher global densities the
user optima are still reached on average. The kind of predictive
information that we implemented depends on the current positions of
all agents in the system. It employs an approximation formula for
the travel times based on these positions. This kind of information
is similar to that used in modern smartphone apps which rely on
crowdsourced
data. The fact that the expected user optima
are realized, also in the Braess states, is a strong hint at the
importance of the paradox in modern traffic networks. As already
proposed by previous observations of real world data
\cite{madrigal2018,cabannes2018impact}, smartphone apps seem to
support the realization of user optima
in road networks.

In scenarios where drivers utilize only their own memories of travel times we
could also show that the user optima of the four test states are
realized. Similar behaviour in a two-route system without
correlations and dynamics similar to TASEP and a similar type of
personal historical information was observed
in~\cite{levy2016emergence}, indicating that this finding is rather general. We
could show that the user optima are also realized in networks with
correlations: both in our 4link and 5link network the expected user
optima are realized. This further strengthens the importance of the
paradox, since the reliance on personal historical information is
likely relevant in many commuter's scenarios.

In future research it would be interesting to consider combinations
of the latter two types of information and see if drivers depending
on different types of information influence each other and the
system as a whole. Are user optima also realized in this case?

\section*{Acknowledgements}
Financial support by Deutsche Forschungsgemeinschaft (DFG), Germany
under grant SCHA 636/8-2 is gratefully acknowledged. Also support by
the Bonn-Cologne Graduate School of Physics and Astronomy (BCGS) is
acknowledged. Monte Carlo simulations were carried out on the CHEOPS
(Cologne High Efficiency Operating Platform for Science) cluster of
the RRZK (University of Cologne).

\appendix
\section{Mixed user optima in Braess' original example}
\label{sec:app_muo}

As described in Sec.~\ref{sec:braess_paradox}, the Braess paradox is
also observed in Braess' original model if users choose their routes
according to mixed strategies. Let $p_{14}$, $p_{23}$ and $p_{153}$
be the probabilities with which all users choose routes 14, 23 and
153, respectively. The probabilities are subject to
$p_{14}+p_{23}=1$ or $p_{14}+p_{23}+p_{153}=1$ for the 4link and
5link systems, respectively.

In the 4link system, for mixed strategies (ms) the expectation values,
denoted by $\langle\, T^{(4)}_{i, \rm ms}\, \rangle$, of the travel times on the
routes 14 and 23 are
\begin{eqnarray}
 \langle\, T^{(4)}_{14, \rm ms}\, \rangle&=  50+(1+p_{14}\cdot(N-1))\cdot 11  \\
 \langle\, T^{(4)}_{23, \rm ms}\, \rangle&=  50+(1+p_{23}\cdot(N-1))\cdot 11
\end{eqnarray}
for each car.

For $N=6$ a mixed user optimum state (muo) is found for
$p_{14}=p_{23}=1/2$ with a travel time expectation value of
$\langle\, T^{(4)}_{14,\rm muo}\, \rangle=\langle\, T^{(4)}_{23,\rm muo}\,
\rangle=88.5$.

In the system with the new road, the expectation values of the travel
times on the three routes are
\begin{eqnarray}
 \langle\, T^{(5)}_{14, \rm ms}\, \rangle&=(1+(p_{14}+p_{153})(N-1))\cdot 10 +50+1+p_{14}(N-1)  \\
  \langle\, T^{(5)}_{23, \rm ms}\, \rangle&=(1+(p_{23}+p_{153})(N-1))\cdot 10 +50+1+p_{23}(N-1)   \\
 \langle\, T^{(5)}_{153, \rm ms}\, \rangle&= (2+(p_{14}+p_{23}+2p_{153})(N-1))\cdot 10 + 10+1+p_{153}(N-1).
\end{eqnarray}
Here a mixed user optimum is given for $p_{14}=p_{23}=5/13$ and
$p_{153}=3/13$ with travel time values
$\langle\, T^{(5)}_{14,\rm muo}\, \rangle=\langle\, T^{(5)}_{23,\rm muo}\,
\rangle=\langle\, T^{(5)}_{153,\rm muo}\, \rangle=93.6923$.

For the case of mixed strategies the expected user optimum travel
times are also higher in the 5link system than in the 4link system,
i.e. the paradox occurs also with mixed strategies.

This example shows that the average number of cars on a specific route
in the mixed user optimum does not have to correspond to the (integer)
number of cars on that route in the pure user optimum: in the 5link
system, the pure user optimum is for $N=6$ given by distributing the
users equally on the three routes. The mixed equilibrium is not
achieved by all users choosing the routes with equal probability!

\section{The test states}\label{sec:app_test_states}

The parameter sets for which the algorithm was tested are marked in
Figs.~\ref{fig:phase_diagrams}~(a) and~(b). The corresponding travel
time values that are expected in the existing pure and mixed user
optima (as found previously in~\cite{bittihn2016}
and~\cite{bittihn2018}) are given in the following.
See~\cite{bittihn2016,bittihn2018} as well for the naming scheme for
the different states.

All these states are boundedly rational user
optima~\cite{mahmassani1987boundedly}:
in~\cite{bittihn2016,bittihn2018} we found the user optima of the
systems by tuning the decisions of the particles externally (either
the $N_{14},N_{23},N_{153}$ for fixed personal strategies or the
$\gamma,\delta$ for turning probabilities) and states for which
$\Delta T=|T_{14}-T_{23}|+|T_{14}-T_{153}|+|T_{23}-T_{153}|<100$
were considered to be user optima. This means that the travel times
on the three routes are not necessarily exactly equal but are
reasonably close to each other and we thus consider the states to be
user optima. Since the travel times are not necessarily exactly
equal, we give here the maximum travel time observed in those states
as the reference time, which is used as a comparison for the system
with intelligently deciding particles.

\paragraph*{state $\star 1$:} has the parameters $L_5=97$ and $M=156$
which correspond to $\hat{L}_{153}/\hat{L}_{14}=0.5$,
$\rho_{\mathrm{global}}^{(4)}\approx 0.13$,
$\rho_{\mathrm{global}}^{(5)}= 0.12$.  For externally tuned particles
this is an ``$E_5$ optimal, all 153" state both for fixed personal
strategies and for turning probabilities.

\noindent In the 4link system
\begin{itemize}
\item a pure user optimum is given for
  $N^{(4)}_{14,\mathrm{puo}}=N^{(4)}_{23,\mathrm{puo}}=78$ which
  corresponds to $n_{\mathrm{l,\,puo}}^{(j_1)(4)}=0.5$ with
  $T_{\mathrm{max,puo}}^{(4)}\approx 692$ and
  $\Delta T^{(4)}_{\mathrm{puo}}\rightarrow 0$.
\item a mixed user optimum is given for
  $\gamma^{(4)}_{\mathrm{muo}}=0.5$ with
  $T_{\mathrm{max,muo}}^{(4)}\approx 693$ and
  $\Delta T^{(4)}_{\mathrm{muo}}\rightarrow 0$.
\end{itemize}
In the 5link system
\begin{itemize}
\item a pure user optimum is given for
  $N^{(5)}_{14,\mathrm{puo}}=N^{(5)}_{23,\mathrm{puo}}=0$ and
  $N^{(5)}_{153,\mathrm{puo}}=156$ which corresponds to
  $n_{\mathrm{l,\,puo}}^{(j_1)(5)}=1.0$ and
  $n_{\mathrm{l,\,puo}}^{(j_2)(5)}=0.0$ with
  $T_{\mathrm{max,puo}}^{(5)}\approx 615$ and
  $\Delta T^{(5)}_{\mathrm{puo}}=0$.
\item a mixed user optimum is given for
  $\gamma^{(5)}_{\mathrm{muo}}=1.0$ and
  $\delta^{(5)}_{\mathrm{muo}}=0.0$ with
  $T_{\mathrm{max,muo}}^{(5)}\approx 615$ and
  $\Delta T^{(5)}_{\mathrm{muo}}=0$.
\end{itemize}

\paragraph*{state $\star 2$:} has the parameters $L_5=339$ and $M=154$
which correspond to $\hat{L}_{153}/\hat{L}_{14}=0.9$,
$\rho_{\mathrm{global}}^{(4)}\approx 0.13$,
$\rho_{\mathrm{global}}^{(5)}= 0.1$.  For externally tuned particles
this is an ``$E_5$ optimal" state both for fixed personal strategies
and for turning probabilities.

\noindent In the 4link system
\begin{itemize}
 \item a pure user optimum is given for $N^{(4)}_{14,\mathrm{puo}}=N^{(4)}_{23,\mathrm{puo}}=77$
 which corresponds to $n_{\mathrm{l,\,puo}}^{(j_1)(4)}=0.5$  with
 $T_{\mathrm{max,puo}}^{(4)}\approx 691$  and $\Delta T^{(4)}_{\mathrm{puo}}\rightarrow 0$.
 \item a mixed user optimum is given for
   $\gamma^{(4)}_{\mathrm{muo}}=0.5$ with
   $T_{\mathrm{max,muo}}^{(4)}\approx 692$ and
   $\Delta T^{(4)}_{\mathrm{muo}}\rightarrow 0$.
\end{itemize}
In the 5link system
\begin{itemize}
\item a pure user optimum is given for $N^{(5)}_{14,\mathrm{puo}}=36$,
  $N^{(5)}_{23,\mathrm{puo}}=30$ and $N^{(5)}_{153,\mathrm{puo}}=88$
  which corresponds to $n_{\mathrm{l,\,puo}}^{(j_1)(5)}\approx 0.8$
  and $n_{\mathrm{l,\,puo}}^{(j_2)(5)}\approx 0.3$ with
  $T^{(5)}_{\mathrm{max,puo}}\approx 670$ and
  $\Delta T^{(5)}_{\mathrm{puo}}\approx 22$.
\item a mixed user optimum is given for
  $\gamma^{(5)}_{\mathrm{muo}}=0.8$ and
  $\delta^{(5)}_{\mathrm{muo}}=0.3$ with
  $T^{(5)}_{\mathrm{max,muo}}\approx 667$ and
  $\Delta T^{(5)}_{\mathrm{muo}}\approx 12$.
\end{itemize}

\paragraph*{state $\star 3$:} has the parameters $L_5=37$ and $M=248$
which correspond to $\hat{L}_{153}/\hat{L}_{14}=0.4$,
$\rho_{\mathrm{global}}^{(4)}\approx 0.21$,
$\rho_{\mathrm{global}}^{(5)}= 0.2$.  For externally tuned particles
this is a ``Braess 1" state both for fixed personal strategies and
for turning probabilities.

\noindent In the 4link system
\begin{itemize}
\item a pure user optimum is given for
  $N^{(4)}_{14,\mathrm{puo}}=N^{(4)}_{23,\mathrm{puo}}=124$ which
  corresponds to $n_{\mathrm{l,\,puo}}^{(j_1)(4)}=0.5$ with
  $T^{(4)}_{\mathrm{max,puo}}\approx 764$ and
  $\Delta T^{(4)}_{\mathrm{puo}}\rightarrow 0$.
 \item a mixed user optimum is given for $\gamma^{(4)}_{\mathrm{muo}}=0.5$
 with $T_{\mathrm{max,muo}}^{(4)}\approx 763$  and $\Delta T^{(4)}_{\mathrm{muo}}\rightarrow 0$.
\end{itemize}
In the 5link system
\begin{itemize}
 \item two pure user optimum exist for
 \begin{enumerate}
 \item $N^{(5)}_{14,\mathrm{puo(i)}}=0$,
   $N^{(5)}_{23,\mathrm{puo(i)}}=124$ and
   $N^{(5)}_{153,\mathrm{puo(i)}}=124$ which corresponds to
   $n_{\mathrm{l,\,puo(i)}}^{(j_1)(5)}=0.5$ and
   $n_{\mathrm{l,\,puo(i)}}^{(j_2)(5)}=0.0$ with
   $T_{\mathrm{max,puo(i)}}^{(5)}\approx 978$ and
   $\Delta T^{(5)}_{\mathrm{puo(i)}}=10$.
 \item $N^{(5)}_{14,\mathrm{puo(ii)}}=124$,
   $N^{(5)}_{23,\mathrm{puo(ii)}}=0$ and
   $N^{(5)}_{153,\mathrm{puo(ii)}}=124$ which corresponds to
   $n_{\mathrm{l,\,puo(ii)}}^{(j_1)(5)}=1.0$ and
   $n_{\mathrm{l,\,puo(ii)}}^{(j_2)(5)}=0.5$ with
   $T_{\mathrm{max,puo(ii)}}^{(5)}\approx 978$ and
   $\Delta T^{(5)}_{\mathrm{puo(ii)}}=11$.
\end{enumerate}
\item a mixed user optimum is given for
  $\gamma^{(5)}_{\mathrm{muo}}=0.87$ and
  $\delta^{(5)}_{\mathrm{muo}}=0.1$ with
  $T_{\mathrm{max,muo}}^{(5)}\approx 895$ and
  $\Delta T^{(5)}_{\mathrm{muo}}\approx 78$.
\end{itemize}

\paragraph*{state $\star 4$:} has the parameters $L_5=218$ and $M=712$
which correspond to $\hat{L}_{153}/\hat{L}_{14}=0.7$,
$\rho_{\mathrm{global}}^{(4)}\approx 0.59$,
$\rho_{\mathrm{global}}^{(5)}= 0.5$.  For externally tuned particles
this is an ``Braess 1" state for fixed personal strategies. For
turning probabilities no user optima could be found for these
parameters.

\noindent In the 4link system
\begin{itemize}
\item a pure user optimum is given for
  $N^{(4)}_{14,\mathrm{puo}}=N^{(4)}_{23,\mathrm{puo}}=356$ which
  corresponds to $n_{\mathrm{l,\,puo}}^{(j_1)(4)}=0.5$ with
  $T_{\mathrm{max,puo}}^{(4)}\approx 1991$ and
  $\Delta T^{(4)}_{\mathrm{puo}}\rightarrow 0$.
 \item there is no short term mixed user optimum due to fluctuating domain walls
\end{itemize}
In the 5link system
\begin{itemize}
\item a pure user optimum is given for
  $N^{(5)}_{14,\mathrm{puo}}=357$, $N^{(5)}_{23,\mathrm{puo}}=178$ and
  $N^{(5)}_{153,\mathrm{puo}}=177$ which corresponds to
  $n_{\mathrm{l,\,puo}}^{(j_1)(5)}=0.75$ and
  $n_{\mathrm{l,\,puo}}^{(j_2)(5)}\approx 0.67$ with
  $T_{\mathrm{max,puo}}^{(5)}\approx 2177$ and
  $\Delta T^{(5)}_{\mathrm{muo}}\approx 24$.
 \item there is no mixed user optimum due to fluctuating domain walls.
\end{itemize}


\section{Accuracy of the travel time predictions used for public predictive information}
\label{sec:how_well_publ_pred_info}

\begin{figure}[ht]
  \centering
  \includegraphics{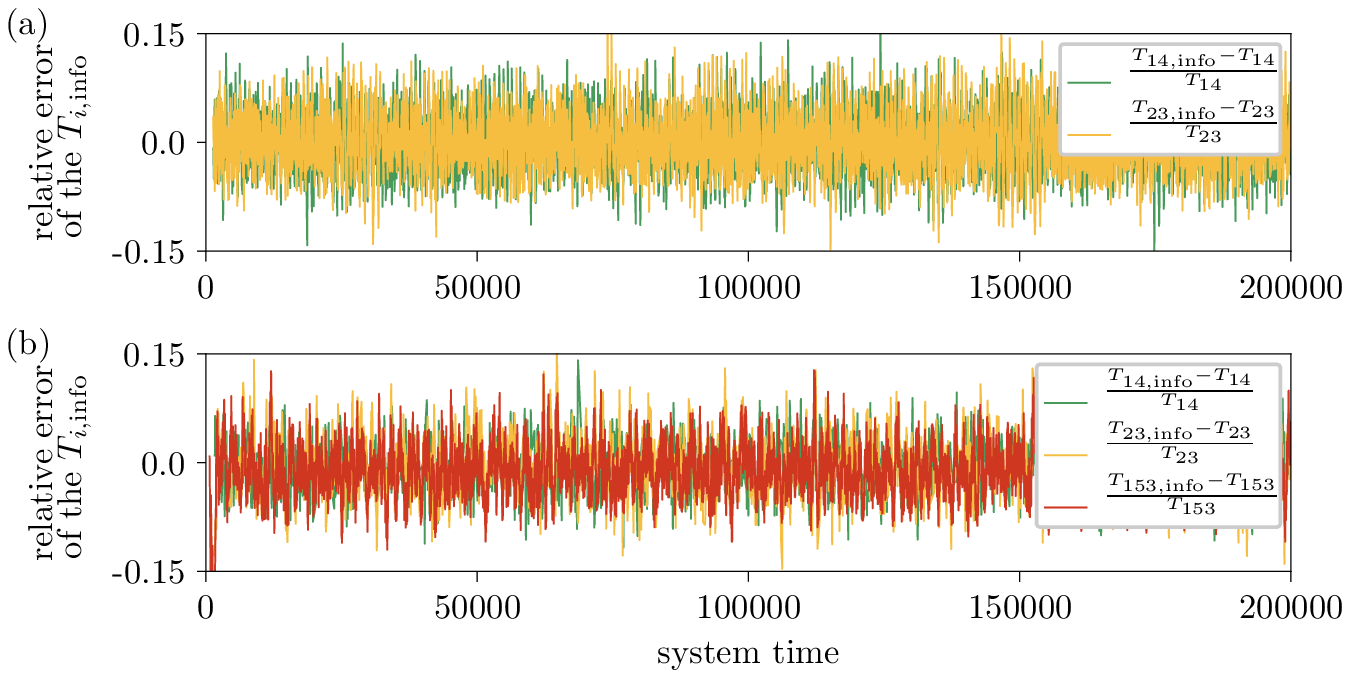}
  \caption{\label{fig:accuracy_state1}Accuracy of the public
    predictive travel time predictions $T_{i,\mathrm{info}}$ for state
    $\star 1$ in (a) the 4link system and (b) the 5link system,
    corresponding to the data shown in
    Figure~\ref{fig:publ_pred_star1}.}
\end{figure}

\begin{figure}[ht]
  \centering
  \includegraphics{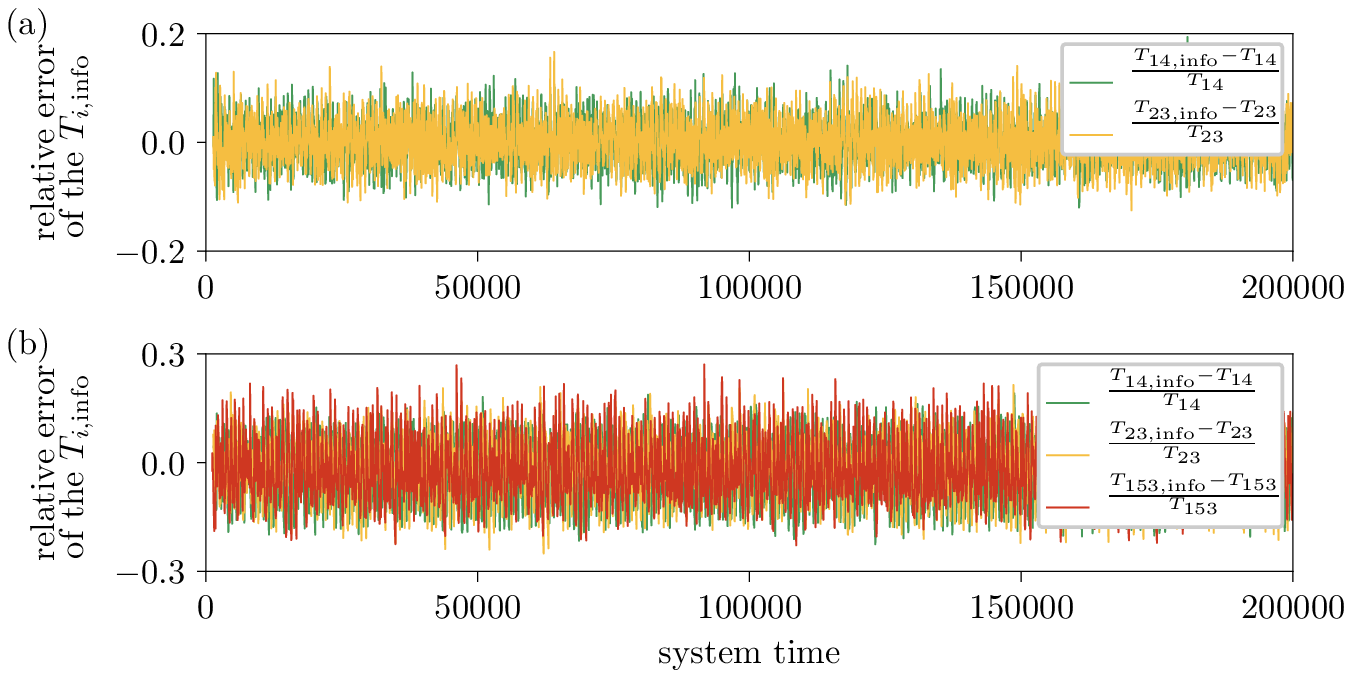}
  \caption{\label{fig:accuracy_state2}Accuracy of the public
    predictive travel time predictions $T_{i,\mathrm{info}}$ for state
    $\star 2$ in (a) the 4link system and (b) the 5link system,
    corresponding to the data shown in
    Figure~\ref{fig:publ_pred_star2}.}
\end{figure}

\begin{figure}[ht]
  \centering
  \includegraphics{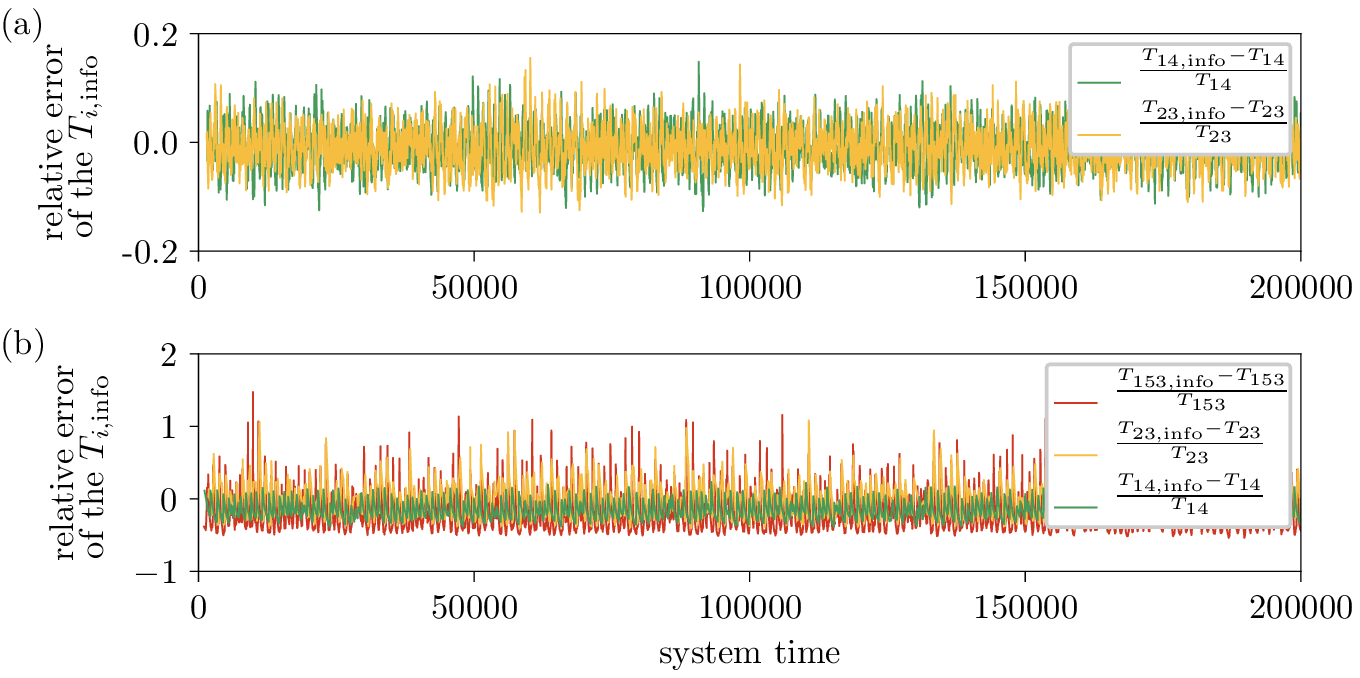}
  \caption{\label{fig:accuracy_state3}Accuracy of the public predictive travel time predictions $T_{i,\mathrm{info}}$ for state $\star 3$ in (a) the 4link system and (b) the 5link system, corresponding to the data shown in Figure~\ref{fig:publ_pred_star3}.}
\end{figure}

\begin{figure}[ht]
  \centering
  \includegraphics{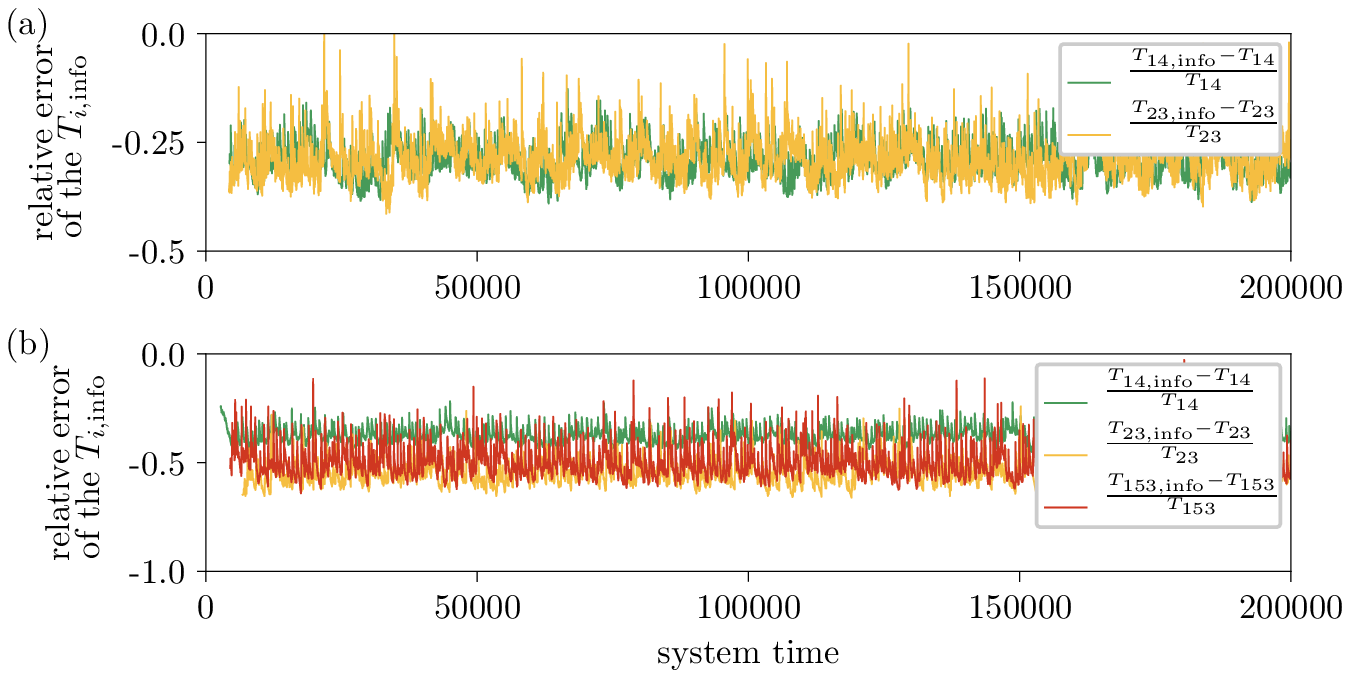}
  \caption{\label{fig:accuracy_state4}Accuracy of the public
    predictive travel time predictions $T_{i,\mathrm{info}}$ for state
    $\star 4$ in (a) the 4link system and (b) the 5link system,
    corresponding to the data shown in
    Fig.~\ref{fig:publ_pred_star4}.}
\end{figure}

In this section of the Appendix we discuss the accuracy of the
travel time predictions that are provided in the case of public
predictive information, i.e. we discuss how accurate the
$T_{i,\mathrm{info}}$ as given by
Eqs.~(\ref{eq:t14_info_publ_pred})--(\ref{eq:t153_info_publ_pred})
are. As can be seen in Figs.~\ref{fig:publ_pred_star1}
to~\ref{fig:publ_pred_star4} and as discussed in
Section~\ref{sec:results}, only in state $\star 1$ the the expected
user optimum is realized in a stable manner, while in states $\star
2$ to $\star 4$ especially in the 5link systems fluctuations
around user optima persist, resulting in (slightly) unequal mean
travel times of the routes. In Figs.~\ref{fig:accuracy_state1}
to~~\ref{fig:accuracy_state4} we show how accurate the travel time
predictions are. For this, the relative error of the
$T_{i,\mathrm{info}}$ compared to the actually measured $T_{i}$ are
given: the $T_{i,\mathrm{info}}$ of all routes $i$, predicted before
starting a new round are saved for each particle. Once an agent
finishes a round, the travel time that the agent actually
experienced on its chosen route $i$ ($T_{i}$) is measured. From this
the relative error is computed as $(T_{i,\mathrm{info}}-T_i)/T_i$.

In Fig.~\ref{fig:accuracy_state1} it can be seen that in state
$\star 1$ the predicted travel times are fairly accurate, as the
relative error of the predictions as compared to the actual measured
travel times lies beneath 15\% at all times. This is not surprising
since the global density is very low in that state and thus only
small jamming effects and other correlation-induced effects are
expected. Judging from Fig.~\ref{fig:publ_pred_star1}~(b) one can
furthermore see that in the 5link system almost all particles use
route 153. Thus the system is more or less a single periodic TASEP,
for which Equation~(\ref{eq:tt_per_tasep}) is correct.

As can be seen in Fig.~\ref{fig:publ_pred_star2}, in state $\star 2$
in the 4link system the user optimum is realized in a stable way.
Judging from Fig.~\ref{fig:accuracy_state2}~(a) we can see that in
this case the travel time predictions are also accurate. This is for
the same reasons as stated above for the 4link system of state
$\star 1$. In the 5link system the user optimum is only reached on
average (Fig.~\ref{fig:publ_pred_star2}). In
Fig.~\ref{fig:accuracy_state2}~(b) it can be seen that the predicted
travel times are off for up to almost $\pm$30\%.

In state $\star 3$, the predictions in the 4link system are still
fairly accurate, as seen in Fig.~\ref{fig:accuracy_state3}~(a).
Accordingly also the user optimum in the 4link is realized well
(Fig.~\ref{fig:publ_pred_star3}). In the 5link system we can see,
that the predicted travel times are of by a large degree, especially
for route 153, they are at times more than 100\% off. In
Fig.~\ref{fig:accuracy_state3}~(b) one can see, that at times when
the travel time on route 153 is predicted too high, the travel time
on route 14 is predicted too low and vice versa. This could be the
cause for the fluctuations around the user optimum that can be seen
in Fig.~\ref{fig:publ_pred_star3}.

In state $\star 4$ the global density has a value of
$\rho_{\mathrm{global}}^{(4)}\approx 0.59$ and
$\rho_{\mathrm{global}}^{(5)}= 0.5$ and is thus much higher than in
the three other states. At a higher global density also jamming
effects, which are not covered by the simple approximations used for
the travel time predictions, become more important. As can be seen
in Fig.~\ref{fig:accuracy_state4}~(a), in the 4link system the travel
times are predicted too low. This could be a consequence of
neglecting correlations and could be the cause of the slightly
higher oscillations around the optimum in the 4link system of state
$\star 4$ (Fig.~\ref{fig:publ_pred_star4}~(b)) as compared to the
4link system of the other states and the resulting higher mean
travel times than those expected from the user optimum in systems
with externally tuned strategies
(Fig.~\ref{fig:publ_pred_star4}~(a)). In
Fig.~\ref{fig:accuracy_state4}~(b) one can see that the travel times
in the 5link system are also predicted too low. This could also be
the reason for the imperfect realization of the 5link's user optimum
(Fig.~\ref{fig:publ_pred_star4}).

\bibliography{braess3}

\end{document}